\documentclass[10pt,onecolumn,superscriptaddress,showpacs,prd,aps,amsmath,amssymb,nofootinbib]{revtex4-2}
\usepackage{graphicx,color,xcolor}
\usepackage{orcidlink}

\usepackage{amsmath,amssymb}
\usepackage{verbatim}

\usepackage{amsfonts}

\usepackage{bm}

\usepackage{hyperref}
\hypersetup{
  colorlinks=true,        
  linkcolor=blue,         
  citecolor=cyan,         
}

\usepackage{cleveref}

\usepackage{derivative}

\usepackage{siunitx}
\DeclareSIUnit{\gauss}{G}
\DeclareSIUnit{\Mass}{\mathit{M}}
\DeclareSIUnit{\c}{\mathit{c}}
\DeclareSIQualifier{\Sun}{\ensuremath{\odot}}

\begin{document}
\title{An efficient approach to resistive GRMHD simulations of binary neutron star mergers}
\author{Miquel Miravet-Tenés\orcidlink{0000-0002-8766-1156}}
\email{miquel.miravet@uv.es}
\affiliation{Mathematical Sciences and STAG Research Centre, University of Southampton, Southampton SO17 1BJ, UK}
\affiliation{Departament d'Astronomia i Astrofísica, Universitat de València, Avinguda Vicent Andrés Estellés 19, 46100, Burjassot (València), Spain}
\author{Ian Hawke\orcidlink{0000-0003-4805-0309}}
\affiliation{Mathematical Sciences and STAG Research Centre, University of Southampton, Southampton SO17 1BJ, UK}
\author{Rahime Matur\orcidlink{0000-0001-9488-3817}}
\affiliation{Mathematical Sciences and STAG Research Centre, University of Southampton, Southampton SO17 1BJ, UK}
\author{Alex J. Wright\orcidlink{0000-0002-5953-4221}}
\affiliation{Mathematical Sciences and STAG Research Centre, University of Southampton, Southampton SO17 1BJ, UK}

\date{\today}

\begin{abstract}
Astrophysical plasmas are often highly conducting, but finite resistivity can nevertheless influence the evolution of strongly magnetised systems. Capturing these effects in general relativistic magnetohydrodynamics is computationally challenging because high conductivity leads to stiff source terms in the evolution equations. Standard numerical approaches rely on implicit-explicit (IMEX) time integration to maintain stability, but this substantially increases the complexity and computational cost of the simulations. Building upon the previously developed Resistive Extension upGrade for Ideal MagnEtohydrodynamics (REGIME), we present a new fluid-frame formulation that simplifies the incorporation of leading-order resistive effects by avoiding matrix inversions required in previous approaches. The method is implemented in a general relativistic magnetohydrodynamics code and validated through one-dimensional current-sheet tests and simulations of isolated magnetised neutron stars and binary neutron star mergers. We find that the new formulation reproduces the expected resistive behaviour with accuracy comparable to IMEX-based methods, while substantially reducing the computational cost and complexity of the simulations. In binary neutron star mergers, finite resistivity leads to measurable differences in the post-merger gravitational-wave signal and remnant structure, showing that resistive effects can produce measurable changes in the post-merger dynamics even in the high-conductivity regime.
\end{abstract}

\maketitle

\section{Introduction}\label{sec::intro}

Binary neutron star mergers provide a unique environment to study the interplay between strong-field gravity, dense nuclear matter, and relativistic magnetised plasmas. The detection of the gravitational-wave (GW) signal associated with GW170817~\cite{gw170817} and its electromagnetic counterpart AT2017gfo~\cite{2017Sci...358.1559K,2017ApJ...848L..12A} demonstrated the importance of understanding the complex post-merger evolution of these systems~\cite{Abbott:2017a,Abbott:2017b,Metzger:2017}. Magnetic fields are expected to play a central role in this evolution, influencing angular momentum redistribution, e.g.~\cite{Kiuchi:2018,Margalit:2022,Radice:2024}, mass ejection and disc formation, e.g.~\cite{Ciolfi:2020,Desai:2023,Most:2023b,Jiang:2025,Gutierrez:2025}, jet launching mechanisms, e.g.~\cite{Nathanail:2021,Combi:2023b,Hayashi:2025}, and the possible production of short gamma-ray bursts, e.g.~\cite{Siegel:2014,Kawamura:2016,Paschalidis:2017}.

Numerical simulations of magnetised compact-object mergers are commonly performed within the framework of ideal general-relativistic magnetohydrodynamics (GRMHD)~\cite{WhiskyMHD,Bucciantini:2011,grhydro,IllinoisGRMHD:2015,spritz,Kiuchi:2022,Neuweiler:2024,Palenzuela:2025,AthenaK:2026}, where the assumption of infinite electrical conductivity leads to the flux-freezing condition, such that magnetic field lines are advected with the fluid. Ideal GRMHD has successfully described many aspects of magnetised neutron star mergers, including magnetic-field amplification through the Kelvin–Helmholtz and magnetorotational instabilities, and the development of large-scale magnetic structures, e.g.~\cite{Price:2006,Siegel:2013,Kiuchi:2015,Kiuchi:2024,Aguilera-Miret:2023,Reboul-Salze:2025}. However, the ideal-MHD approximation prevents changes in magnetic topology and neglects non-ideal processes associated with finite conductivity.

In realistic astrophysical plasmas, the electrical conductivity is large but finite~\cite{Harutyunyan:2016,Harutyunyan:2018}, and many astrophysical processes cannot be fully described within ideal MHD because they involve departures from the flux-freezing approximation~\cite{Goedbloed_Poedts_2004,Yamada:2010}. Finite resistivity introduces physical processes absent in ideal MHD, including magnetic diffusion, Ohmic dissipation, and magnetic reconnection. By relaxing the flux-freezing condition, resistivity allows magnetic field lines to slip relative to the plasma, enabling changes in magnetic topology and the redistribution of magnetic flux. These processes can modify magnetic-field evolution, magnetic-energy dissipation and conversion, angular momentum transport, and the coupling between the electromagnetic field and the fluid~\cite{Biskamp:1996,Fleming:2000,Sano:2002,Pessah:2008}. Although many astrophysical plasmas are expected to remain close to the ideal-MHD regime, even small deviations from ideal behaviour can influence the evolution of strongly magnetised systems. Numerical studies including finite resistivity have explored a variety of systems, including rotating neutron stars~\cite{Cheong:2025,Franceschetti:2025}, binary neutron star mergers~\cite{Dionysopoulou:2015,BAMpaper}, magnetorotational instabilities~\cite{Qian:2017}, magnetic reconnection~\cite{Lyubarsky:2005,Mignone:2012,Ripperda:2019a,Sironi:2025}, and jet launching from accreting black holes~\cite{Qian:2018,DelZanna:2023,Mattia:2024,Nathanail:2025}. However, despite their physical importance, simulations that explicitly include finite resistivity remain relatively limited due to the additional numerical complexity associated with evolving non-ideal effects, particularly in the high-conductivity regime relevant to many astrophysical applications.

These numerical difficulties are particularly severe in the context of numerical relativity. Unlike ideal GRMHD, where the electric field is constrained by the ideal Ohm's law, resistive GRMHD promotes the electric field to an independent dynamical variable and evolves it according to the generalized Ohm's law. The corresponding evolution equations contain stiff source terms associated with the rapid relaxation towards the ideal-MHD limit at high conductivity. Explicit time integration methods then suffer from severe timestep restrictions because the resistive relaxation timescale becomes much shorter than the dynamical timescale in the high-conductivity regime. To overcome this difficulty, current resistive GRMHD simulations generally employ implicit-explicit (IMEX) time integration schemes\footnote{Although there also exist previously developed methods~\cite{Komissarov:2007, Takamoto:2011} and more recent approaches~\cite{Cordero-Carrion:2025}.}, where the non-stiff transport terms are treated explicitly and the stiff resistive terms implicitly~\cite{Ascher:1995,Pareschi:2005}. The first application of these methods in special-relativistic resistive MHD was done in Ref.~\cite{Palenzuela:2009}. Building on top of this framework, the IMEX method was extended to resistive GRMHD simulations in~\cite{Bucciantini:2013} and~\cite{Dionysopoulou:2013}, implementing the full resistive form of Ohm's law. More recently, IMEX schemes have been implemented in several different codes with the aim of simulating neutron stars and their mergers~\cite{Cheong:2022,Cheong:2025,Dionysopoulou:2015,Franceschetti:2025}, and, they have also been implemented in GPU-accelerated numerical relativity codes~\cite{Shankar:2023,Azizi:2026}. These methods provide a robust treatment of resistive effects across a wide range of conductivities, but their implementation requires substantial modifications to existing GRMHD codes, including the evolution of additional dynamical variables~\cite{Bugli:2014,DelZanna:2016}, modifications of primitive-variable recovery procedures~\cite{Ripperda:2019b}, and the introduction of new time integrators~\cite{Palenzuela:2013,Qian:2017,Cheong:2022,Franceschetti:2025}.

An alternative approach is to exploit the fact that many astrophysical systems of interest are expected to operate in the high-conductivity regime, where deviations from ideal MHD remain small. In this regime, resistive effects can be treated perturbatively as corrections around the ideal-MHD equilibrium state. The original idea of Ref.~\cite{jinRelaxationSchemesSystems1995} demonstrates, using the Chapman-Enskog theory~\cite{Chapman:1970,Cercignani:2002}, that a system of coupled equations with potentially stiff terms can be reduced to a modified system, given a certain limit. This is the logic we follow in the present work, and it has been already implemented in special-relativistic MHD~\cite{Wright:2020,Hatton:2024} and limited general-relativistic cases~\cite{Wright_thesis}. This approach captures the leading-order resistive corrections without requiring the implicit treatment of stiff source terms, providing a computationally efficient alternative to full resistive GRMHD.

In this work, we present the general-relativistic extension of this perturbative resistive framework, the Resistive Extension upGrade for Ideal MagnEtohydrodynamics (REGIME). This approach introduces finite-resistivity corrections through an expansion around ideal GRMHD and modifies the evolution equation of the vector potential by adding the leading-order resistive contribution. To this end, we employ fluid-frame variables, from which the Eulerian electric-field correction can be recovered through simple algebraic relations, avoiding the matrix inversions required by the Eulerian-variable formulation of~\cite{Wright_thesis}. The REGIME formulation avoids the need for expensive IMEX schemes and allows resistive effects to be incorporated into existing ideal-GRMHD infrastructures with minimal changes. The method is particularly suited for astrophysical scenarios where the conductivity is high and resistive effects act as small corrections to the ideal-MHD evolution. Furthermore, unlike phenomenological approaches, this framework is derived from first principles and therefore does not require parameter tuning for different astrophysical scenarios.

This work is intended as a proof-of-principle study presenting the new REGIME formulation based on the fluid-frame observer. We implement REGIME within the \texttt{IllinoisGRMHD} code~\cite{IllinoisGRMHD:2015, 2023PhRvD.107d4037W, 2026PhRvD.113d3045C}, part of the Einstein Toolkit infrastructure~\cite{Loffler:2012, 2025zndo..15520463R}, and validate the method through a series of increasingly complex tests, including evolutions of one-dimensional resistive current sheets~\cite{Komissarov:2007} and magnetised Tolman-Oppenheimer-Volkoff (TOV)~\cite{Tolman:1939,Oppenheimer:1939} stars. These tests demonstrate the ability of REGIME to reproduce the expected resistive behaviour while preserving the computational advantages of explicit ideal-GRMHD approaches. Finally, we explore the impact of finite resistivity on the GW signal and post-merger evolution of a magnetised neutron star binary, finding measurable differences in the compactness of the merger remnant and magnetic-field strength and structure compared to the ideal-MHD case.

The paper is laid out as follows. In~\cref{sec::formalism} we present the REGIME formalism with the help of a couple of examples, and develop, step by step, the fluid-frame approach. We describe the numerical setup within the \texttt{IllinoisGRMHD} code in~\cref{sec::num_setup}, and we test this implementation by evolving self-similar current sheets and magnetised TOV stars in~\cref{sec::tests}. In~\cref{sec::bns_simulations}, we perform and analyse simulations of binary neutron star mergers with explicit resistivity. Finally, in~\cref{sec::conclusions}, we summarise the findings of the previous sections, discuss the insights that our results provide, and propose future directions. Unless otherwise stated, equations and quantities are written in geometric units. Spacetime indices are denoted by Latin letters $a$,...,$f$ and run from 0 to 3, while spatial indices are denoted by Latin letters $i$,...,$l$ and run from 1 to 3.

\section{Formalism}\label{sec::formalism}

REGIME provides a framework for approximating resistive magnetohydrodynamics, as well as other near-ideal stiff systems, through a perturbative expansion around the ideal limit. Although this procedure may, in principle, be implemented algorithmically (see in particular~\cite{Wright:2020,Hatton:2024}), the resulting sequence of operations is generally laborious and, for systems of even moderate complexity, is typically tractable only with the aid of computer algebra or numerical matrix inversion.

The present work investigates whether this procedure may be simplified through the exploitation of underlying symmetries, together with, where appropriate, the introduction of further controlled approximations.

\subsection{The Jin-Xin Example}
\label{sec:jinxin}

The standard example~\cite{jinRelaxationSchemesSystems1995} is the stiff system
\begin{subequations}
\label{eq:example}
    \begin{align}
        \label{eq:example1}
        \pdv{u}{t} + \pdv{v}{x} &= 0 \, , \\
        \label{eq:example2}
        \pdv{v}{t} + a \pdv{u}{x} &= \epsilon^{-1} \left[ f(u) - v \right] \, .
    \end{align}
\end{subequations}
In the limit $\epsilon \to 0$ the system is stiff, and $v \to f(u)$.

In the stiff limit the full system from~\cref{eq:example} is equivalent to a single scalar nonlinear conservation law. The parameter $\epsilon$ is linked to the relaxation timescale of the system towards this reduced single variable model. When the timescale is vanishingly small the reduced single variable model is a good approximation. The interesting case is where this relaxation timescale is small but not negligible. In that case, we can hope to construct a reduced single-variable model by a perturbation analysis about the singular limit $\epsilon \to 0$. We therefore assume that $v = f(u) + \epsilon v_1$. If the initial data is \emph{well-prepared} then this is a reasonable assumption, as the variables will be driven to this regime on a timescale $\mathcal{O}(\epsilon)$. With this perturbation assumption we see that \cref{eq:example2} becomes 
\begin{equation}
    \label{eq:example3}
    \pdv{f(u)}{t} + a \pdv{u}{x} = v_1 + \mathcal{O}(\epsilon) \, .
\end{equation}
Using the chain rule and the equation of motion for $u$, \cref{eq:example1} we can write
\begin{equation}
    \label{eq:example4}
    v_1 = \left[ a - (f')^2 \right] \pdv{u}{x} + \mathcal{O}(\epsilon) \, .
\end{equation}
Finally, we can go back to the equation of motion for $u$, \cref{eq:example1}, to get
\begin{equation}
    \label{eq:example5}
    \pdv{u}{t} + \pdv{f(u)}{x} = \epsilon \pdv{}{x} \left\{ \left[ (f')^2 - a \right] \pdv{u}{x} \right\} \, ,
\end{equation}
where we have dropped terms $\mathcal{O}(\epsilon^2)$.

The single equation of motion~\cref{eq:example5} is the reduced model which should capture the leading order behaviour near the singular limit $\epsilon \to 0$.

\subsection{Newtonian MHD example}

In the following, we explore how this perturbative approximation works in Newtonian resistive MHD. This provides a simple example in which the expansion can be applied explicitly to the electromagnetic evolution equations, allowing us to identify the leading-order resistive correction. As in the Jin--Xin example discussed in~\cref{sec:jinxin}, we expand the system around its equilibrium state and derive a reduced evolution equation that remains well behaved in the ideal-MHD limit.

The standard formulation of \emph{Newtonian} electromagnetism consists of the evolution equations for the electric and magnetic fields:
\begin{subequations}
    \label{eq:newt_maxwell}
    \begin{align}
        \label{eq:newt_maxwell_B}
        \pdv{\vec B}{t} + \nabla \times \vec E &= 0 \,, \\
        \label{eq:newt_maxwell_E}
        \pdv{\vec E}{t} - \nabla \times \vec B &= \epsilon^{-1} \vec J \, .
    \end{align}
\end{subequations}
In the ideal MHD limit we expect the ``resistivity'', represented by $\epsilon$, to behave as $\epsilon \to 0$, which drives $\vec J \to \vec E + \vec v \times \vec B$. We apply the same expansion approach as in the Jin-Xin example of~\cref{sec:jinxin} to this system. First, we assume that
\begin{equation}
    \label{eq:newt_maxwell_ce1}
   \vec E = - \vec v \times\vec B + \epsilon \vec E^{(1)} \, .
\end{equation}
Plugging this into the evolution equation for the electric field, \cref{eq:newt_maxwell_E}, gives
\begin{equation}
    \label{eq:newt_maxwell_ce2}
    \vec E^{(1)} = - \pdv{(\vec v \times \vec B)}{t} - \nabla \times \vec B + \mathcal{O}(\epsilon) \, .
\end{equation}
For now we will assume that the time derivative term can be neglected.
Using the near-equilibrium approximation from~\cref{eq:newt_maxwell_ce1} together with the explicit correction form in~\cref{eq:newt_maxwell_ce2} to leading order in the evolution equation for the magnetic field,~\cref{eq:newt_maxwell_B}, we find
\begin{equation}
\label{eq:newt_maxwell_ce3}
    \pdv{\vec B}{t} - \nabla \times \left(\vec
    v \times \vec B \right) = -\epsilon \nabla \times \nabla \times \vec B \, .
\end{equation}
Assuming that the velocity $\vec v$ is known independently of the electric field, this is a reduced model for the magnetic field evolution. It limits to standard ideal MHD as $\epsilon \to 0$ and is not stiff in that limit.

Using a standard vector identity, combined with the divergence free constraint $\nabla \cdot \vec B = 0$, we can rewrite the reduced model for the magnetic field as
\begin{equation}
\label{eq:newt_maxwell_ce4}
    \pdv{\vec B}{t} - \nabla \times \left( \vec v \times \vec B \right) = \epsilon \nabla^2 \vec B \, .
\end{equation}
This makes the correction term away from the MHD limit explicitly look like a dissipation term.

Remember that we neglected the time derivative of the leading order magnetic field. In general, as employed in~\cite{Wright:2020,Wright_thesis}, we can use the equations of motion to re-write the time derivatives in terms of spatial derivatives, leading to more complex additional terms. In addition, as the velocity appears in the time derivatives this will couple the resistive source term to the fluid explicitly. However, we will see later that it is possible to bypass this issue by working in the fluid frame.

\subsection{Electromagnetism}

We are interested in (resistive) GRMHD using a vector potential approach. This is widely used in simulating neutron star mergers~\cite{IllinoisGRMHD:2015,spritz}. The vector potential approach ensures the divergence-free condition for the magnetic field. We will focus, for now, on the electromagnetic aspects. The notation here is chosen to match (mostly) that used in the \texttt{IllinoisGRMHD} code~\cite{IllinoisGRMHD:2015}.

Considering a vector potential $\mathcal{A}_a$, one can build the Faraday tensor using the exterior derivative of the vector potential, $F_{ab} = \nabla_{[a} \mathcal{A}_{b]}$. The Maxwell equations are then
\begin{subequations}
    \label{eq:maxwell}
    \begin{align}
        \label{eq:maxwell1}
        \nabla_a F^{ab} &= \mathcal{J}^b \, , \\
        \label{eq:maxwell2}
        \nabla_a F^{* ab} &= 0 \, ,
    \end{align}
\end{subequations}
where $\mathcal{J}^b$ is the charge density current and $F^{* ab}$ is the dual of the Faraday tensor.

The derivation of the reduced-order model is typically carried out at the level of the decomposed equations of motion. To this end, we introduce a 3+1 foliation defined by the unit timelike vector normal to the spatial hypersurfaces, $n^a$. The electric and magnetic fields measured by the Eulerian observer (with 4-velocity $n^a$) are then defined as follows:
\begin{subequations}
    \label{eq:spatial_eb}
    \begin{align}
        \label{eq:spatial_e}
        E_b &= n^a F_{ab} \, , \\
        \label{eq:spatial_b}
        B_b &= n^a F^*_{ab} = \frac{1}{2} n^a \varepsilon_{abcd} F^{cd} \, 
    \end{align}
\end{subequations}
where we introduce the volume element $\varepsilon_{abcd} = [abcd]/\sqrt{-g}$, with $[abcd]$ being the space-time Levi-Civita pseudo-tensor. Note that we will later write the lapse $\alpha$ and shift $\beta^i$, which follow from the normal to the foliation as $n^a = \alpha^{-1} ( 1, -\beta^i)$. We can also project the charge current along the normal observer as
\begin{equation}
    \label{eq:spatial_j_split}
    \mathcal{J}^b = q n^b + J^b \, .
\end{equation}
The projected vector potential along the Eulerian observer is
\begin{equation}
    \label{eq:spatial_A_split}
    \mathcal{A}^b = \Phi n^b + A^b \, .
\end{equation}
The fluid 4-velocity is \emph{not} projected but written as $u^0 (1, v_{(n)}^i)$. This implies that
\begin{equation}
    \label{eq:spatial_u_split}
    u^a = \alpha u^0(n^a+v_{(n)}^a) =       u^0 \left( \alpha n^a + \left[ v^a + \beta^a \right] \right) \, .
\end{equation}
Combining these all together, the scalar isotropic resistive closure is written in terms of Eulerian variables as~\cite{Anile:1989}
\begin{equation}
    \label{eq:resistive_defn_spatial}
    J^b = q v_{(n)}^b + \eta^{-1} \Gamma \left[ E^b + \varepsilon^{bcde} u_c B_d n_e - E^c v^{(n)}_c v_{(n)}^b \right] \, ,
\end{equation}
where $\eta$ stands for the scalar, isotropic resistivity, and $\Gamma = \alpha u^0$, for the Lorentz factor. We see how this closure will lead to a stiff source term in the Maxwell equation~\cref{eq:maxwell1}. We note that the above expression may also be written in terms of the fluid three-velocity, $v^i$.

To complete this system we need to write out the Maxwell equations as split equations of motion. For our purposes we need the evolution equations for the vector potential and for the electric field. These are
\begin{subequations}
    \label{eq:eom}
    \begin{align}
        \label{eq:eom1}
        \partial_t A_i &= -\alpha E_i - \varepsilon_{ijk} \beta^j B^k - \partial_i \left( \alpha \Phi - \beta^j A_j \right) \, , \\
        \label{eq:eom2}
        \partial_t E^i &= \gamma^{-1/2} \partial_j \left[ \alpha \gamma^{1/2} (\gamma^{il} \gamma^{jm} - \gamma^{im} \gamma^{jl}) \partial_l A_m \right] \\
        & \quad - 4 \pi \alpha J^i + \alpha K E^i + \mathcal{L}_\beta E^i \, . \nonumber 
    \end{align}
\end{subequations}
We have split the normal vector $n^a$ as
\begin{equation}
    \label{eq:lapse_shift}
    n^a = \alpha^{-1} \begin{pmatrix} 1, - \beta^i \end{pmatrix} 
\end{equation}
and introduced the 3-metric $\gamma^{ij}$ and its determinant $\gamma$, the trace of the extrinsic curvature $K$, and the induced volume element $\varepsilon_{ijk} = n^d \varepsilon_{ijkd}$.

We note that the equations of motion~\cref{eq:eom} are not complete. First, the time component of the vector potential, $\Phi$, is not given. That is set by a gauge choice. Second, the spatial charge current $J^i$ needs imposing. This is set by the choice of physical system; in our case, this is scalar isotropic resistivity given by~\cref{eq:resistive_defn_spatial}. Thirdly, the coupling to the fluid or plasma is defined through the 4-velocity, which is assumed to be known (although in reality the electromagnetic stress tensor and the perfect fluid stress tensor are coupled through the conservation of stress-energy). Finally, the full Maxwell equations~\cref{eq:maxwell} contain additional constraints.

For definiteness, we therefore write out the full system as
\begin{subequations}
    \label{eq:full_system}
    \begin{align}
        \label{eq:full_system1}
        \partial_t A_i &= -\alpha E_i - \varepsilon_{ijk} \beta^j B^k - \partial_i \left( \alpha \Phi - \beta^j A_j \right) \, , \\
        \label{eq:full_system2}
        \partial_t E^i &= \gamma^{-1/2} \partial_j \left[ \alpha \gamma^{1/2} (\gamma^{il} \gamma^{jm} - \gamma^{im} \gamma^{jl}) \partial_l A_m \right] \\
        & \quad -4 \pi \alpha J^i + \alpha K E^i + \mathcal{L}_\beta E^i \, , \nonumber \\
        \label{eq:full_system3}
        J^i &= q v_{(n)}^i + \eta^{-1} \Gamma \left[ E^i + \varepsilon^{ijk} v^{(n)}_j B_k - E^j v^{(n)}_j v_{(n)}^i \right] \, , \\
        \label{eq:full_system4}
        \partial_i \gamma^{1/2} E^i &= \gamma^{1/2} q \, ,
    \end{align}
\end{subequations}
We note that the ``no-monopole'' constraint $\partial_i \gamma^{1/2} B^i = 0$ follows automatically from the definition of the magnetic field in terms of the vector potential. We also note that the ``Ohm's law''~\cref{eq:full_system3} is the closure for scalar isotropic resistivity from~\cref{eq:resistive_defn_spatial} using our final notation.

Note that in addition we have the orthogonality conditions,
\begin{subequations}
    \label{eq:orthog}
    \begin{align}
        n^i A_i &= 0 \, , \\
        n^i E_i &= 0 \, , \\
        n^i B_i &= 0 \, , 
    \end{align}
\end{subequations}
which should hold automatically, but need not be guaranteed to hold for the perturbative expansion based models.

\subsection{REGIME}
\label{sec:regime_eqns}

\subsubsection{Steps}\label{sec:D.1}

As the evolution equation for the electric field~\cref{eq:full_system2} contains the charge current as a source, and the closure relation for the charge current~\cref{eq:full_system3} contains a term that is stiff in the ideal limit where $\eta \to 0$, we see that the reduced model we seek will eliminate the equation of motion for the electric field, and constrain this quantity. The resulting perturbation expansion that allows for appropriate behaviour in the ideal limit is
\begin{equation}
    \label{eq:regime1}
    E^i = -\varepsilon^{ijk}{v_{(n)}}_j B_k + \eta E^i_{(1)} = -\varepsilon^{ijk}\alpha^{-1} (v_j+\beta_j) B_k + \eta E^i_{(1)} \,,
\end{equation}
assuming that the resistivity $\eta$ introduces a small correction to the ideal limit (i.e $\eta \ll 1$). To leading order, this is the standard form of Ohm's law for ideal MHD, as we should expect. We see that our reduced model is given by
\begin{equation}
    \label{eq:regime_reduced}
    \partial_t A_i = \varepsilon_{ijk}v^j B^k - \partial_i \left( \alpha \Phi - \beta^j A_j \right) - \eta \alpha E_i^{(1)} \, .
\end{equation}
This is precisely the ideal MHD equation with an algebraic correction given by the first order correction term $E^i_{(1)}$, which needs to be found. We only need to construct the MHD system of equations to leading order in $\eta$. We therefore evaluate the full system~\cref{eq:full_system} at $\mathcal{O}(\eta^0)$, using the short-hand $E^i_{(0)} = -\varepsilon^{ijk} {v_{(n)}}_j B_k$, finding
\begin{subequations}
    \label{eq:full_system_lo}
    \begin{align}
        \label{eq:full_system_lo1}
        \partial_t A_i &= -\alpha \varepsilon^{ijk} v_j B_k - \partial_i \left( \alpha \Phi - \beta^j A_j \right) \, , \\
        \label{eq:full_system_lo2}
        \partial_t E^i_{(0)} &= \gamma^{-1/2} \partial_j \left[ \alpha \gamma^{1/2} (\gamma^{il} \gamma^{jm} - \gamma^{im} \gamma^{jl}) \partial_l A_m \right] - 4 \pi \alpha (q v_{(n)}^i + \Gamma E^i_{(1)}) + \alpha K E^i_{(0)} + \mathcal{L}_\beta E^i_{(0)} \, , \\
        \label{eq:full_system_lo4}
        \partial_i \gamma^{1/2} E^i_{(0)} &= \gamma^{1/2} q \, .
    \end{align}
\end{subequations}
We have directly substituted the appropriate part of the Ohm's law, \cref{eq:full_system3}, using the perturbation expansion~\cref{eq:regime1}, into the evolution equation for the electric field~\cref{eq:full_system2}, in order to get~\cref{eq:full_system_lo2}.

The leading order term for the electric field evolution, \cref{eq:full_system_lo2}, is the definition of the first order correction we seek, $E^i_{(1)}$. The other terms in the equation are either spatial derivatives of known quantities (everything on the right-hand-side, including the charge density $q$, which is found using~\cref{eq:full_system_lo4}), or time derivatives of known quantities (the left hand side). These quantities are known because the leading order term $E^i_{(0)}$ is given from the velocity and magnetic field, and the magnetic field is given by the vector potential.

To compute the first order correction $E^i_{(1)}$, we need to re-write the time derivatives of the leading order electric field. This can be computed, and~\cite{Wright_thesis} shows the steps required. However, this requires the inversion of multiple matrices, and required constraints (such as~\cref{eq:orthog}) are not guaranteed to hold, potentially causing numerical issues. Therefore, a simpler alternative would be preferred.

\subsubsection{Fluid frame version}

In the form of the Maxwell equations used here, the coupling between the Faraday tensor (and hence the electric and magnetic fields) and a fluid or plasma is through the 4-velocity of the particles, denoted $u^a$. We can then introduce the electric and magnetic fields as measured by the fluid observer as
\begin{subequations}
    \label{eq:fluid_eb}
    \begin{align}
        \label{eq:fluid_e}
        e_b &= u^a F_{ab} \, , \\
        \label{eq:fluid_b}
        b_b &= u^a {F^*}_{ab} = \frac{1}{2} u^a \varepsilon_{abcd} F^{cd} \, .
    \end{align}
\end{subequations}
 We can also project the charge current along the fluid observer as
\begin{equation}
    \label{eq:fluid_j_split}
    \mathcal{J}^b = q_0 u^b + j^b \, .
\end{equation}
As an alternative to the Eulerian observer presented in~\cref{sec:D.1}, let us use the fluid frame. Use the closure condition~\cref{eq:resistive_defn_spatial} to write, now in terms of the fluid observer~\cite{Bucciantini:2013},
\begin{equation}
    \label{eq:resistive_defn_fluid}
    j^b = \eta^{-1} e^b \, .
\end{equation}
We note that in the ideal limit the fluid frame electric field vanishes, $e_c^{(0)} = 0$. We therefore expand the fluid electric field as
\begin{equation}
    \label{eq:regime_fluid1}
    e_c = 0 + \eta e_c^{(1)}
\end{equation}
from which we have $j_c = e_c^{(1)}$. We then write out the Maxwell equation~\cref{eq:maxwell1} as 
\begin{equation}
    \label{eq:regime_fluid_reduced}
    q_0 u^a + e^a_{(1)} = \nabla_b \left[ \varepsilon^{abcd} b_c u_d \right] + \mathcal{O}(\eta) \, .
\end{equation}
We immediately get, to leading order, 
\begin{equation}
    \label{eq:regime_fluid_e1}
    e^f_{(1)} = \perp^f_a \nabla_b \left[ \varepsilon^{abcd} b_c u_d \right] \, .
\end{equation}
Here, $\perp^f_a = \delta^f_a + u^f u_a$ is the projector with respect to the fluid observer. The right-hand side contains time derivatives that we wish to eliminate. Using the metric-compatible connection, $\nabla_e \varepsilon^{abcd} = 0$, we decompose $\nabla_b$ into a spatial part and a time derivative along the normal observer worldlines:
\begin{equation}
    \nabla_b = D_b - n_b n^e\nabla_e\,,
\end{equation}
which yields
\begin{equation}
    e^f_{(1)} = \perp^f_a D_b\left[\varepsilon^{abcd}b_c u_d\right] + V^f\,,
\end{equation}
where $V^f$ collects all terms involving time derivatives. Expanding explicitly:
\begin{align}\label{eq:timeder_term}
    V^f & = -\perp^f_a n_b n^e\nabla_e\left[ \varepsilon^{abcd}b_c u_d\right] \\
    & = -\perp^f_a\left(n^e\nabla_e\left[\varepsilon^{abcd}n_b b_c u_d\right] - \varepsilon^{abcd}b_c u_d \, n^e\nabla_e n_b\right) \nonumber \\
    & = -\perp^f_a \left(n^e\nabla_e\left[\varepsilon^{abcd}n_b b_c u_d\right]-\varepsilon^{abcd}a_b b_c u_d\right)\,, \nonumber 
\end{align}
where we have used the Leibniz rule and identified the acceleration vector of the normal observers, $a_b = n^e \nabla_e n_b$, which is related to the lapse function by:
\begin{equation}
    a_b = \nabla_b \ln \alpha\,.
\end{equation}
The two terms in the last line of \cref{eq:timeder_term} have distinct physical interpretations. The first is the time derivative, along the normal observer worldlines, of the spatial cross product of the magnetic and velocity fields. The second is the triple cross product of the normal observer acceleration $a_b$ with $b_c$ and $u_d$. Both terms can be neglected under conditions appropriate to a binary neutron star merger remnant: the magnetic field and fluid velocity evolve on timescales much longer than the dynamical timescale, making the first term small in the quasi-static approximation; the lapse gradient $a_b = \nabla_b \ln\alpha$, while not small in magnitude, is directed roughly along the same radial direction as both $b_c$ and $u_d$ in the remnant, so the triple cross product is suppressed by the near-coplanarity of these three vectors. Under these assumptions, the leading-order correction to the fluid electric field reduces to the computation of purely spatial derivatives:
\begin{equation}\label{eq:regime_fluid_e1_spatial}
    e^f_{(1)} \approx \perp^f_a D_b\left[\varepsilon^{abcd}b_c u_d\right]\,.
\end{equation}

\subsubsection{From the fluid frame to the Eulerian frame}
\label{sec:REGIME_fluid_to_eulerian}

All current numerical implementations use a spatial (i.e. Eulerian) frame for the evolution (e.g.~\cite{IllinoisGRMHD:2015,Neuweiler:2024,Kiuchi:2022,AthenaK:2026}). Therefore, we need to convert the correction term from the fluid frame into the spatial frame.

The Faraday tensor is
\begin{subequations}
    \begin{align}
    F^{ab} & =  U^aE_{U}^b-U^bE_U^a+\varepsilon^{abcd}U_cB_d^U\,,  \\
    F^{*ab} & = U^aB_{U}^b-U^bB_U^a-\varepsilon^{abcd}U_cE_d^U\,, \label{eq:dual_daraday_gen_obs}
\end{align}
\end{subequations}
where $U^a$ is a generic observer, and $E_U^a$ and $B_U^a$ are the electric and magnetic fields measured by this observer, respectively. As shown in~\cref{eq:fluid_eb}, the electric and magnetic fields can be obtained by contracting the Faraday tensor with the observer. We can decompose the observer as
\begin{equation}
    U^a = \alpha u^0 (n^a+\alpha^{-1}[v^a+\beta^a])\,,
\end{equation}
keeping in mind that $n^a = \alpha^{-1} (1,-\beta^i)$. The Eulerian observer, with 4-velocity $n^a$, will measure the following electric and magnetic fields:
\begin{subequations}
\begin{align}
    E^a & = F^{ab} n_b = \alpha u^0E_{U}^a+U^aE_U^bn_b+n_b\varepsilon^{abcd}u^0(v_c+\beta_c)B_d^U \\
    B^a & = {F^{*ab}} n_b = \alpha u^0 B_{U}^a+U^aB_U^bn_b-n_b\varepsilon^{abcd}u^0(v_c+\beta_c)E_d^U\,. 
\end{align}
\end{subequations}
If we write the RHS of the previous equations in terms of the fluid observer, we will have
\begin{subequations}
\begin{align}
    E^a & = \alpha u^0 e^a+u^ae^bn_b+u^0\varepsilon^{abcd}n_b(v_c+\beta_c)b_d \label{eq:eul_electric}\\
    B^a & = \alpha u^0 b^a-\alpha b^0 u^a-u^0\varepsilon^{abcd}n_b(v_c+\beta_c)e_d\,. 
\end{align}
\end{subequations}
The last term in \cref{eq:eul_electric} is the only one that appears in the ideal case, as $e^a = 0$ (see \cref{eq:regime1}). Therefore, the first order correction to the electric field in the Eulerian frame, which needs to be included in \cref{eq:regime_reduced}, is
\begin{equation}\label{eq:1st_ord_eul_Efield}
    E^a_{(1)} = \alpha u^0 e^a_{(1)} + e^b_{(1)} n_b u^a\, .
\end{equation}
We need to compute the comoving velocity $u^a$ and magnetic field $b^a$ from \cref{eq:regime_fluid_e1_spatial} in terms of the fields in the spatial frame. The fluid velocity $u^a$ can easily be expressed in terms of the fluid 3-velocity using \cref{eq:spatial_u_split}. We can contract the dual Faraday tensor from \cref{eq:dual_daraday_gen_obs} with the fluid observer to get an expression of $b^a$ in terms of the Eulerian variables:
\begin{align}
    b^a &= {F^{*ab}}u_b \nonumber \\
    &= n^au^0B^i(v_i+\beta_i)-\alpha u^0B^a -n_c\varepsilon^{abcd}u^0(v_b+\beta_b)E_d\, ,
\end{align}
and one can see that the last term does not necessarily vanish in the non-ideal case. This would introduce a first-order correction to the fluid magnetic field. To avoid this issue, we will consider the fluid magnetic field to leading order in~\cref{eq:eul_electric}. The $0^{th}$ component of the comoving magnetic field will be
\begin{equation}
    b^0 = \frac{u^0}{\alpha}B^i(v_i+\beta_i)\,,
\end{equation}
and it does not involve any electric field, even in the non-ideal case. The spatial component will be obtained from $b_a = F^*_{ab}u^b$. In the ideal MHD limit (i.e.\ to leading order), $b^i$ takes the form~\cite{Anton:2006}
\begin{equation}
    b^i = \frac{1}{\alpha}\Big[\frac{B^i}{u^0}+u^0B^j(v_j+\beta_j)v^i\Big]\,.
\end{equation} 

This approach, which employs the fluid frame, offers several advantages over the one presented in \cite{Wright_thesis}. To obtain the first-order correction to the Eulerian electric field, only a simple algebraic relation such as \cref{eq:1st_ord_eul_Efield} is required. Likewise, the fluid electric field contribution (which already constitutes the first-order correction) only requires knowledge of the fluid velocity and magnetic fields, which can also be related to their Eulerian counterparts through simple algebraic relations. Therefore, there is no need to compute matrix inversions, which may be ill-conditioned. Secondly, the orthogonality conditions in~\cref{eq:orthog} hold automatically, whereas in the original approach they can be violated, potentially leading to uncontrolled errors. This easier implementation makes the fluid-frame approach more portable between different GRMHD numerical codes. 

However, this new approach presents its own caveats. First, for simplicity, we neglect the time derivatives arising in \cref{eq:regime_fluid_e1}. Second, as mentioned above, we neglect the first-order correction to the fluid magnetic field. Including this additional contribution would make the calculation of the fluid electric field considerably more complicated. Thus, the Eulerian approach presented in \cite{Wright_thesis} appears to be more complete, although it faces greater numerical difficulties and may not satisfy the orthogonality conditions.

\section{Numerical setup}\label{sec::num_setup}

\subsection{Evolution of the GRMHD equations}

We perform the evolution of the Einstein equations coupled with the MHD equations, in a dynamical and curved spacetime, with the \texttt{IllinoisGRMHD} code~\cite{IllinoisGRMHD:2015, 2023PhRvD.107d4037W}, built on the \texttt{GRHayL} library~\cite{Cupp:2026}, and part of the Einstein Toolkit infrastructure~\cite{Loffler:2012, 2025zndo..15520463R}. The simulations employ the BSSN formulation~\cite{Baumgarte:1998,Shibata:1995} to evolve the spacetime metric, which is provided by the \texttt{Baikal} thorn, generated with the NRPy+ infrastructure~\cite{Etienne:2018}. It adopts the standard moving-puncture gauge conditions, which include the ``1+log'' slicing condition for the lapse and the hyperbolic gamma-driver condition for the shift. To mitigate high-frequency modes that may arise during the evolution, we apply fifth-order Kreiss-Oliger dissipation, with
a strength of 0.2.

To compute the fluxes of the GRMHD equations, \texttt{IllinoisGRMHD} uses a second-order finite-volume high-resolution shock-capturing (HRSC) scheme (see Sect.~3 from~\cite{IllinoisGRMHD:2015}). Spatial reconstruction is performed via the Piecewise Parabolic Method (PPM)~\cite{Colella:1984}, and the Riemann problem is solved with the standard Harten–Lax–van Leer (HLL) approximate solver~\cite{HLL:1983}. To maintain numerical stability in regions of very low density, a background atmosphere is introduced with a baryonic rest-mass density fixed at a certain floor value, which will vary with the system studied, and will be specified in the examples below.

As previously stated in~\cref{sec::formalism}, the magnetic sector is evolved through the four-vector potential ($\mathcal{A}_a$) rather than the magnetic field itself, which automatically preserves the divergence-free condition for the magnetic field. To mitigate the generation of unphysical magnetic field components, \texttt{IllinoisGRMHD} employs the generalized Lorenz gauge proposed in~\cite{Farris:2012}. At the outer boundaries, outflow conditions are enforced for both the primitive variables and the vector potential~\cite{IllinoisGRMHD:2015}. 

The time evolution is performed within the method-of-lines framework using a 4th-order Runge-Kutta time integrator and a Courant–Friedrichs–Lewy (CFL) coefficient of 0.25 (unless otherwise stated).

\subsection{REGIME implementation} \label{sec:regime_implementation}

We implement the resistive REGIME expansion to leading order within the \texttt{IllinoisGRMHD} code. Below we summarize the key algorithmic steps. 

\subsubsection{Evolution equation}

The vector potential is evolved according to~\cref{eq:regime_reduced}. The \texttt{IllinoisGRMHD} code employs a Yee staggered lattice in which $A_x$, $A_y$, and $A_z$ are located at edge-centered positions $(i,j+\tfrac{1}{2},k+\tfrac{1}{2})$, $(i+\tfrac{1}{2},j,k+\tfrac{1}{2})$, and $(i+\tfrac{1}{2},j+\tfrac{1}{2},k)$, respectively. This staggering ensures that discrete derivatives of $\bf{A}$ automatically satisfy $\nabla\cdot\bf{B}=0$.

\subsubsection{Electric field computation}

The first-order Eulerian electric field $E^a_{(1)}$ is computed directly at edge locations. 

First, the three-velocity and all metric quantities are interpolated from cell centres to cell faces, where the spatial derivatives are computed. The magnetic field is natively defined at face centres: at each face, the normal component is used directly, while the transverse components are interpolated from the neighbouring faces. All interpolations are second-order accurate and consist of two-point averaging along each direction in which the variable is offset from the target location. From these, the $\mathcal{F}$ tensor is computed at faces using
\begin{equation}
    \label{eq:impln_Fab}
    \mathcal{F}^{ab} = \varepsilon^{abcd} b_c u_d  .
\end{equation}
This is actually the Faraday tensor in ideal MHD. Next, the partial derivatives of the $\mathcal{F}$ tensor are computed at edges, using
\begin{equation}
    \label{eq:impln_DF}
    D_j \mathcal{F}^{aj} = \frac{1}{\sqrt{\gamma}}\partial_j (\sqrt{\gamma}\mathcal{F}^{aj})  +\Gamma^a_{jc}\mathcal{F}^{cj}\,,
\end{equation}
and the last term vanishes, since $\mathcal{F}^{ab}$ is antisymmetric. All derivatives of the $\mathcal{F}$ tensor are computed using second order finite differencing.

The remaining steps are purely algebraic. First, our approximation of the correction to the electric field in the fluid frame $e_{(1)}^a$, \cref{eq:regime_fluid_e1_spatial}, is given by the hypersurface projection of the covariant derivatives of the electromagnetic tensor computed in~\cref{eq:impln_DF}. Finally, we need to convert back to the spatial frame as discussed in~\cref{eq:1st_ord_eul_Efield}, giving
\begin{equation}
    \label{eq:eulerian_efield2}
    E^a_{(1)} = \alpha u^0 e_{(1)}^a + (e_{(1)}^b n_b)\,u^a.
\end{equation}
The quantities required to complete the computation of $E^a_{(1)}$ (the 4-velocity $u^a$ and the normal vector $n^a$) have been interpolated to the cell edges.
With these steps we have the first order correction to the electric field in the Eulerian frame $E^a_{(1)}$. We can therefore correct the evolution equation for the vector potential, which is evaluated at the edge locations where we have computed the update term $E^a_{(1)}$ using~\cref{eq:regime_reduced}:
\begin{equation}
    \label{eq:regime_reduced_repeated}
    \partial_t A_i = \varepsilon_{ijk}v^j B^k - \partial_i \left( \alpha \Phi - \beta^j A_j \right) - \eta \alpha E_i^{(1)} \, .
\end{equation}
The only additional pieces needed to construct the correction, which is the final term in this equation, are the lapse $\alpha$ (which is again interpolated to edges) and the resistivity $\eta$.

\subsubsection{Resistivity profile}

A noteworthy feature of our resistivity implementation in~\cref{eq:regime_reduced_repeated} is that the resistivity itself is included purely algebraically in the updates. Therefore a nonlinear (data-dependent) or spatially varying and dynamical resistivity profile is trivial to employ. This is perfectly possible within, for example, methods that rely on IMEX schemes~\cite{Dionysopoulou:2013,Franceschetti:2025,Azizi:2026}, although the varying stiffness of the update terms can make load balancing and efficiency more complex. Including a nonlinear or spatially varying resistivity profile is much more involved in the original REGIME approach based around the equations of motion as constructed in~\cite{Wright_thesis}.

In this proof-of-principle work we choose a simple resistivity profile that varies spatially as a function of baryon density:
\begin{equation}
\eta(\rho) = \frac{\eta_0-\eta_f}{2}\left[1+\tanh\!\left(\frac{\log_{10}(\rho/\rho_c)}{\lambda}\right)\right],
\label{eq:eta_tanh}
\end{equation}
where $\eta_0$ is the peak resistivity, $\eta_f \ll \eta_0$ is the lowest resistivity value, $\rho_c$ is the critical transition density, and $\lambda$ is the logarithmic width. This smoothly interpolates between (near) ideal MHD ($\rho\ll\rho_c$) and fully resistive ($\rho\gg\rho_c$) regimes. Setting $\rho_c=0$ recovers a uniform resistivity $\eta=\eta_0$. All three parameters are runtime-steerable. The purpose of this profile is to minimize the impact of our modifications near the surface and in the exterior artificial atmosphere. The impact of this choice will be discussed further in~\cref{sec:TOV}.

\section{Tests}\label{sec::tests}

The tests performed here are to demonstrate that the model and implementation are ``sufficiently correct'' for the applications of interest, which are neutron star merger simulations. As our approach here is a numerical implementation of a model that approximates the near-ideal asymptotic limit of resistive GRMHD, we should expect residual ``model error'' even in the continuum limit. We first show, via simple tests in flat space with an exact solution, that this model error is small. Then we extend our tests to curved spacetimes and show that the method works efficiently for more interesting cases.

\subsection{Self-similar current sheet}

The self-similar current sheet is a common test, proposed in~\cite{Komissarov:2007} and employed to evaluate magnetic diffusion in a plasma with finite resistivity.
All variables except one are constant and $v^i = 0$, resulting in a system which is initially in static equilibrium. The only non-constant quantity is
\begin{equation}
    \label{eq:sscs1}
    B_y(x, t) = B_0 \mathrm{erf} \left( \frac{x \sqrt{\sigma}}{2 \sqrt{t}} \right) \, ,
\end{equation}
where $\sigma = \eta^{-1}$ is the conductivity and $\mathrm{erf}$ is the error function. This component of the magnetic field then follows a simple diffusion process, as
\begin{equation}
    \label{eq:sscs2}
    \partial_t B_y = \eta\partial^2_xB_y\,.
\end{equation}
This corresponds to the vector potential
\begin{equation}
    \label{eq:sscs3}
    A_z(x, t) = -B_0 \left[ x \, \mathrm{erf}(\theta) + 2 \sqrt{\frac{t}{\pi \sigma}} \exp(-\theta^2) \right] \, ,
\end{equation}
where we have used the shorthand $\theta = \tfrac{x}{2}\sqrt{\tfrac{\sigma}{t}}$. It is important to note that this analytical solution is only valid for infinite pressure~\cite{Bucciantini:2013}. Therefore, instead of using the exact solution for computing the error norm later, we will follow the same procedure as in~\cite{Bucciantini:2013,Cheong:2022,BAMpaper} by using a high-resolution run as our reference solution.

The specific values used in our tests are
\begin{equation}
    \label{eq:sscs4}
    \begin{pmatrix}
        \rho, P, B_0
    \end{pmatrix} = 
    \begin{pmatrix}
        1, 5000, 1
    \end{pmatrix} \,,
\end{equation}
with the fluid velocity initially being zero.
When $\eta = \num{e-3}$ in geometric units (which corresponds to \qty{5e-9}{\s}), and the domain is $x \in [-1.5, 1.5]$ a solution starting from $t=1$ spreads across the domain by $t \sim 100$. Note that for practical implementation the solution uses a shifted time to avoid $t = 0$. We use an ideal fluid with $\Gamma = 4/3$. We expect a sign reversal of $B_y$ across a thin current layer with a certain width. The resistivity is set to $\eta = \num{e-3}$ and $\eta = \num{e-2}$, and we employ different resolutions: $\Delta x = \{1/50,1/100,1/200,1/400\}$. Our reference run has a resolution of $\Delta x = 1/800$. For these simple one-dimensional tests, the spacetime metric is flat, and the CFL factor is set to 0.5.

\begin{figure*}[htbp]
    \centering
    \includegraphics[width=0.8\textwidth]{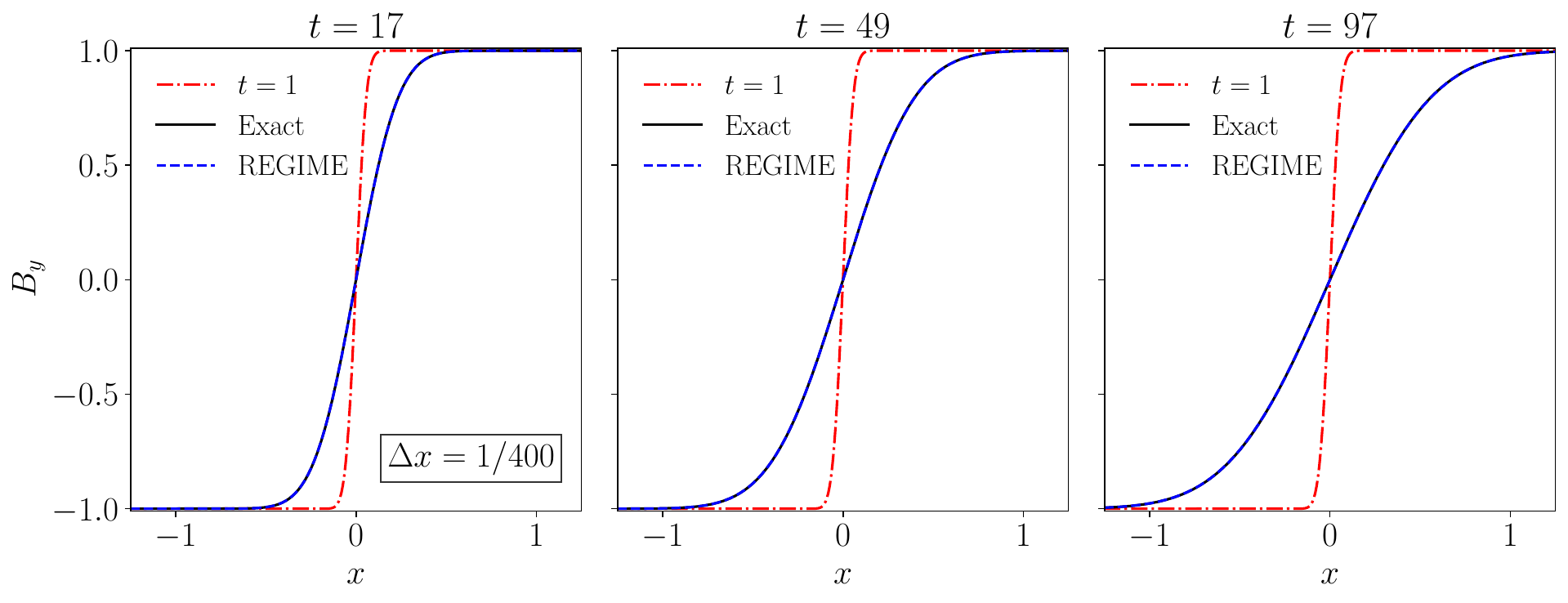}
    \caption{Profile of $B_y$ along the $x$ direction. The black solid line depicts the analytical solution given in~\cref{eq:sscs1}, and the blue dashed line, the result from REGIME. The dash-dotted curve shows the solution at $t = 1$, the initial time.  Each column stands for a different time: $t=\{17, 49, 97\}$. For this case, the resolution is $\Delta x = 1/400$ in the $x$ direction.}
    \label{fig:current_sheet_N1200}
\end{figure*}

In~\cref{fig:current_sheet_N1200} we depict the profile of $B_y$ along the $x$ direction (top row), at three different times: $t=\{17, 49, 97\}$. We compare the exact solution (black solid curve) with the REGIME prediction (blue dashed curve), and we show the REGIME solution at the initial time, $t = 1$. We use the run with resolution $\Delta x = 1/400$. As the system evolves, the width of the current layer significantly increases. By looking at the good agreement between our result and the analytical one, we can conclude that the magnetic field evolution is well described by the REGIME approach.  

\begin{figure*}[htbp]
    \centering
    \includegraphics[width=\textwidth]{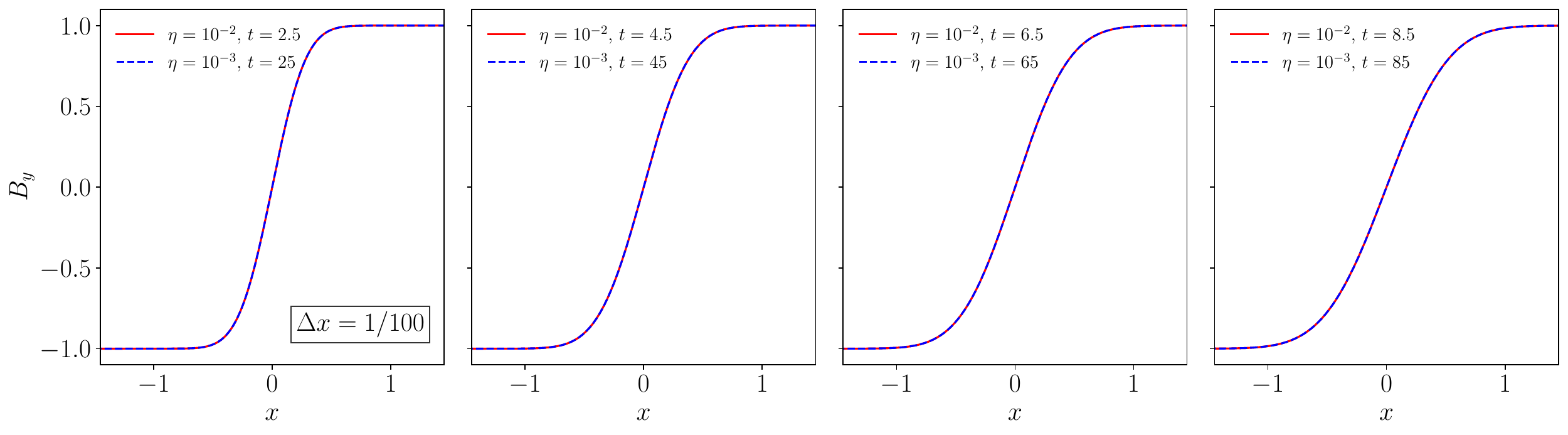}
    \caption{Evolution of the current sheet solution for $B_y$, for different resistivities and at different times. Red solid lines represent solutions with $\eta = \num{e-2}$, whereas the blue dashed lines depict solutions with $\eta = \num{e-3}$. Each panels stands for a different time, which is scaled by the resistivity in order to match the solutions with different resistivity values.}
    \label{fig:current_sheet_eta_comp}
\end{figure*}

As shown in~\cref{eq:sscs1}, we expect the diffusion timescale to scale linearly with the resistivity. A higher resistivity (i.e lower conductivity) will lead to a faster diffusion. In~\cref{fig:current_sheet_eta_comp}, we compare two cases with different resistivities (differing by a factor 10), at different times. The snapshots of the lower resistivity case with $\eta = \num{e-3}$ are taken at a time 10 times larger than the one considered for $\eta = \num{e-2}$. As expected, the profiles coincide. 

\begin{figure*}[htbp]
    \centering
    \includegraphics[width=0.4\textwidth]{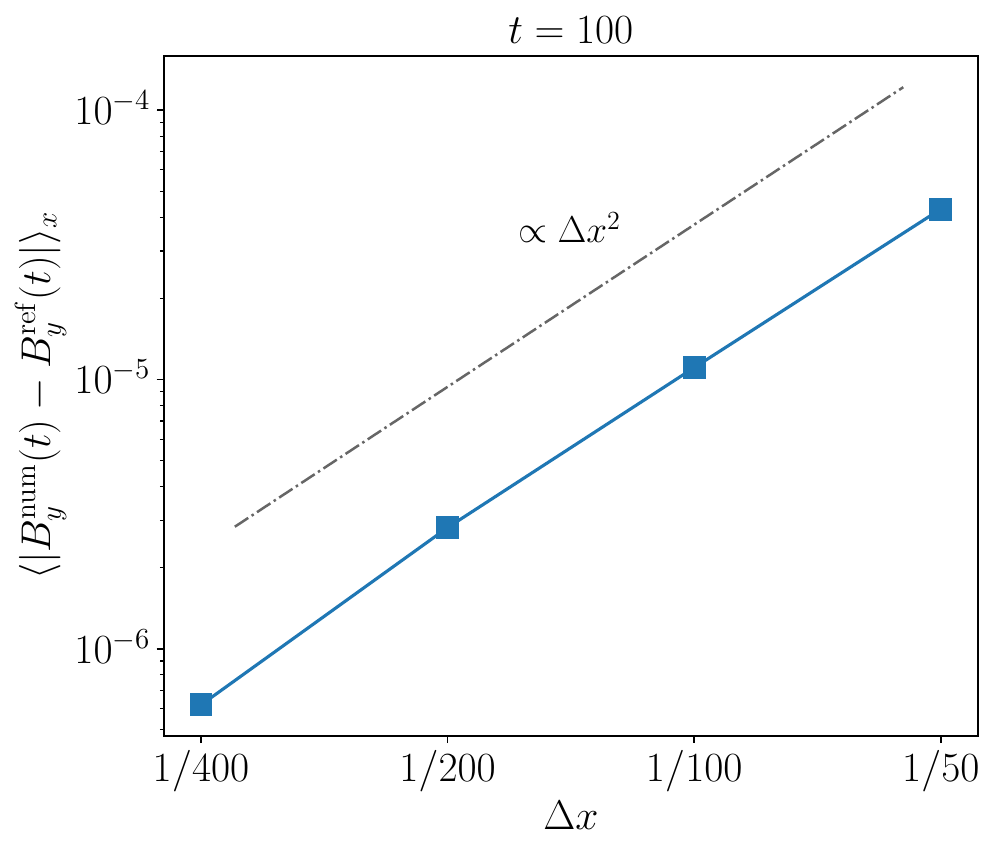}
    \caption{Values of the $L_1$ error norm (blue squares), as a function of the resolution employed in the current sheet tests. The grey dash-dotted line shows second-order convergence.}
    \label{fig:current_sheet_L1}
\end{figure*}

In~\cref{fig:current_sheet_L1} we depict the scaling of the $L_1$ error norm:
\begin{equation}
    L_1(t, \Delta x) = \langle |B_y^{\mathrm{num}}(t) - B_y^{\mathrm{ref}}(t)| \rangle_x\,,
\end{equation}
where we compute the difference between the numerical solution of $B_y$ and the solution obtained with the reference run, and perform an average over space. In order to study the scaling of the $L_1$ norm with resolution, we select a fixed time, e.g. $t=100$. It can be seen that the error scales as $\propto\Delta x^2$, showing second-order convergence. This is in agreement with other works that implement IMEX schemes~\cite{Bucciantini:2013,Cheong:2022,BAMpaper}, where the same order of convergence is found.

\subsection{Magnetised TOV star}
\label{sec:TOV}

In this Section, we show the results of simulations of a stationary, spherically symmetric star given by the TOV equation~\cite{Tolman:1939, Oppenheimer:1939} with the inclusion of an initially purely poloidal magnetic field confined inside the star. The only non-zero component of the vector potential (in a spherical coordinate basis) is $A_\phi$, which is set as
\begin{equation}
    \label{eq:Aphi}
    A_\phi = \varpi^2 A_b\mathrm{max}(P-P_\mathrm{cut},0)^{n_s}\,,
\end{equation}
where $\varpi$ is the cylindrical radius, $P_\mathrm{cut}$ corresponds to the cut-off pressure and is set to $4\%$ of the maximum pressure, and $A_b$ indicates the initial magnetic field strength. To be consistent with other works, e.g.~\cite{Dionysopoulou:2013,Franceschetti:2025,Azizi:2026}, we set $n_s = 2$. The Cartesian components turn out to be
\begin{subequations}
    \label{eq:Avec_tov}
    \begin{align}
     A_x &= -y A_b \mathrm{max}(P-P_\mathrm{cut},0)^2\,, \\
     A_y &= x A_b \mathrm{max}(P-P_\mathrm{cut},0)^2\,,\\
     A_z &= 0\,.
    \end{align}
\end{subequations}
For these simulations, we set $A_b = \num{3.31e3}$ and $P_\mathrm{cut} = \num{6.53e-6}$ in geometric units, resulting in an initial central magnetic field $B_\mathrm{c} \approx  \qty{e15}{\gauss}$. The initial hydrodynamical data is created with the \texttt{TOVSolver} thorn. We build a TOV star with mass $\qty{1.4}{\Mass\Sun}$ and a radius of $\qty{12}{\km}$, with a polytropic equation of state with $K = 100$ and $\Gamma = 2$, and a central density $\rho_c(0) = \qty{7.9e14}{\g\per\cm\cubed}$. The atmosphere is set by a baryonic mass density of $\rho_\mathrm{atm} = \qty{7.981e7}{\g\per\cm\cubed}$. We employ a baseline resolution on the finest grid of $\Delta x = 0.2$ (equivalent to $\qty{295}{\m}$), with 4 refinement levels, and a CFL factor of 0.25. For convergence tests, we also employ a finer resolution of 0.167 and a coarser resolution of 0.3.

The resistivity follows the profile shown in~\cref{eq:eta_tanh}. We choose $\eta_f = 0$, resulting in an ideal MHD exterior. Other works on resistive GRMHD assume vacuum at the stellar exterior~\cite{Dionysopoulou:2013,Franceschetti:2025,Azizi:2026}. This simplification ensures that the conductivity is zero in the stellar atmosphere, which makes the IMEX scheme numerically stable. The REGIME approach, on the contrary, naturally reduces to the ideal case when $\eta \rightarrow 0$, since it is directly proportional to the resistivity. We believe that the ideal MHD approach for the stellar exterior is much more realistic than just considering the vacuum equations, since the magnetosphere still contains (low-density) highly-magnetised plasma~\cite{Goldreich:1969,Lehner:2012,Wright_thesis}. 

\begin{figure*}[htbp]
    \centering
    \includegraphics[width=\textwidth]{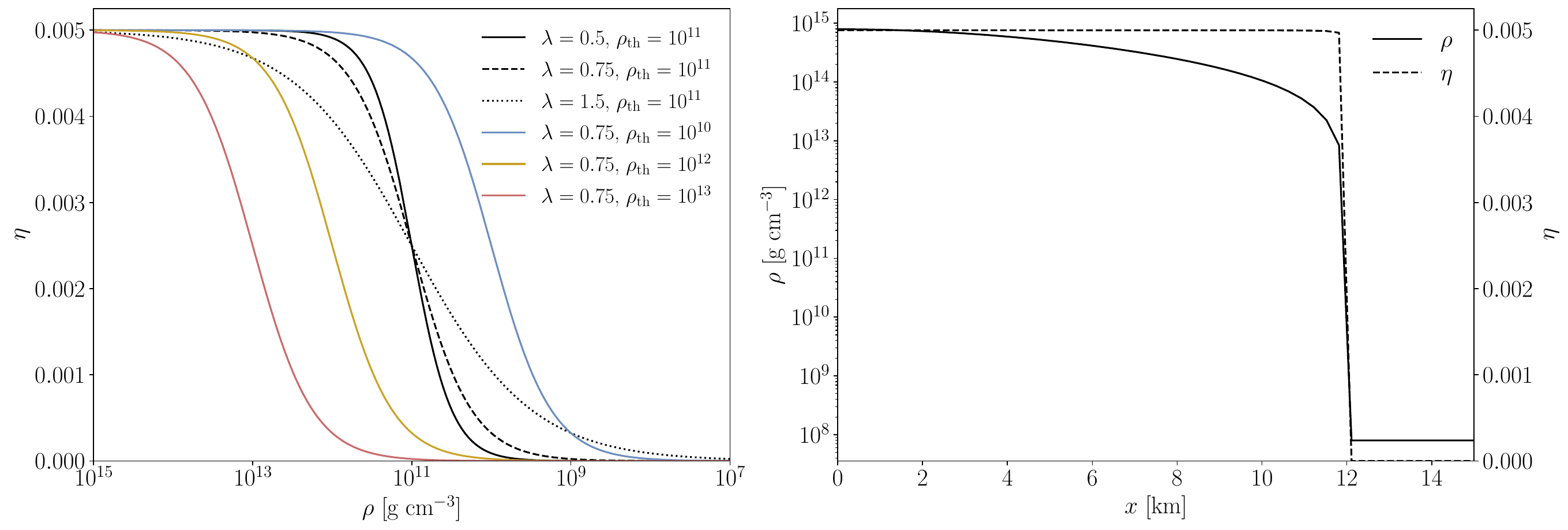}
    \caption{Left panel: Profiles of resistivity as a function of mass density. These curves follow~\cref{eq:eta_tanh}. Different line styles depict different values of $\lambda$, and different colours stand for different values of $\rho_\mathrm{th}$. In all cases, $\eta_f$ is fixed to zero. Right panel: Density (solid curve) and resistivity (dashed curves) profiles as a function of the x-coordinate at the equator of the TOV star, at $t = 0$, for $\rho_\mathrm{th} =10^3\rho_\mathrm{atm}$ and $\lambda = 0.75$. }
    \label{fig:eta_profile}
\end{figure*}

\Cref{fig:eta_profile} shows how $\eta$ depends on the mass density $\rho$. The left panel depicts the relation given in~\cref{eq:eta_tanh} for different values of $\lambda$ and $\rho_\mathrm{th}$, and the right panel shows the radial profile of the mass density and the resistivity (computed using the previous expression) of the TOV star with a polytropic EOS, at $t=0$ and fixing $\rho_{\rm th} = 10^3$ and $\lambda = 0.75$.

\begin{figure}[htbp]
    \centering
    \includegraphics[width=0.5\linewidth]{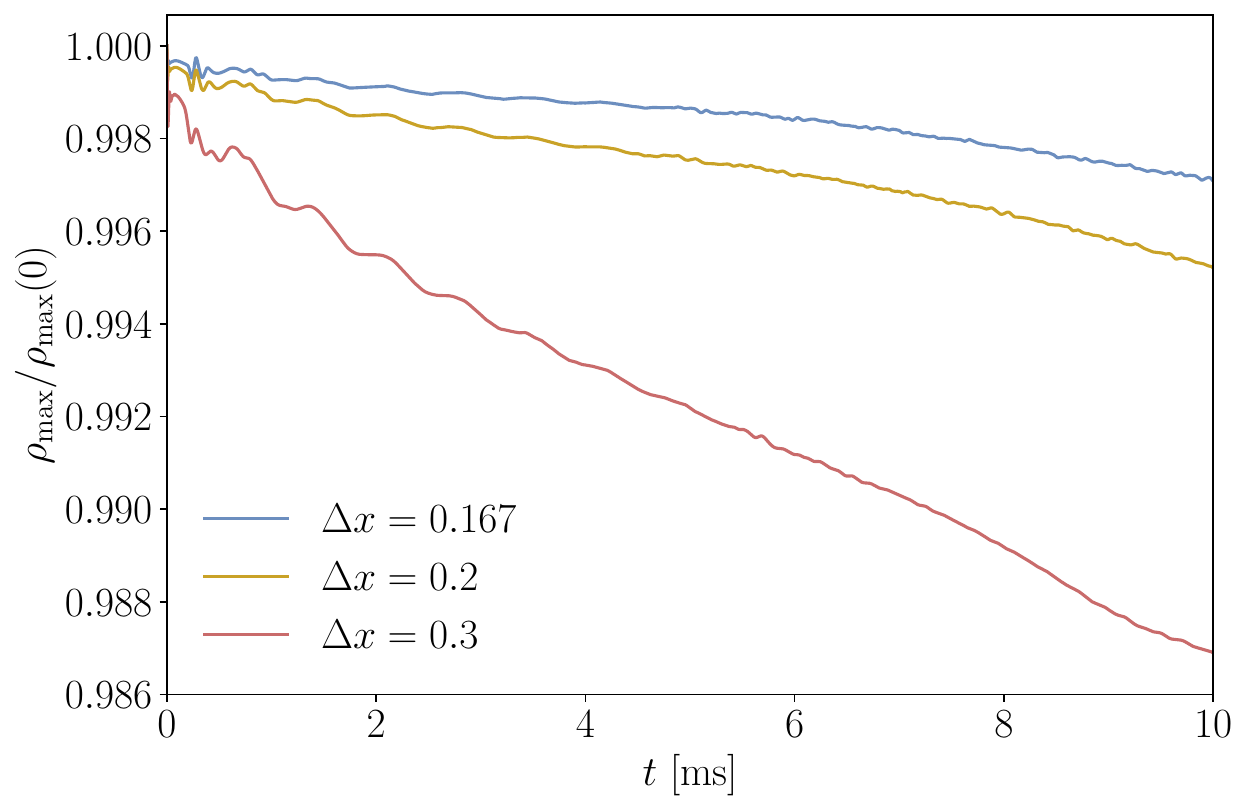}
    \caption{Time evolution of the maximum value of the baryonic mass density, normalised by its initial value, for three different resolutions: $\Delta x= \{0.167,0.2,0.3\}$. After \qty{10}{\ms}, the maximum density value has decreased less than 2\% for the lowest resolution case.}
    \label{fig:rho_max_tev}
\end{figure}

We have performed several simulations with different values of $\eta_0$, $\rho_\mathrm{th} = 10^3\rho_\mathrm{atm}$ and $\lambda = 0.75$. \Cref{fig:rho_max_tev} depicts the time evolution of the maximum baryonic mass density normalised by its initial value, for different resolutions and $\eta =0.0$. One can observe that, after \qty{10}{\ms}, the central density is still larger than the 98.5\% of its initial value, for all resolutions. This behaviour is consistent with previous IMEX simulations in the near-ideal limit, while the present formulation achieves this without requiring the additional complexity of an IMEX implementation.

\begin{figure}[htbp]
    \centering
    \includegraphics[width=0.5\linewidth]{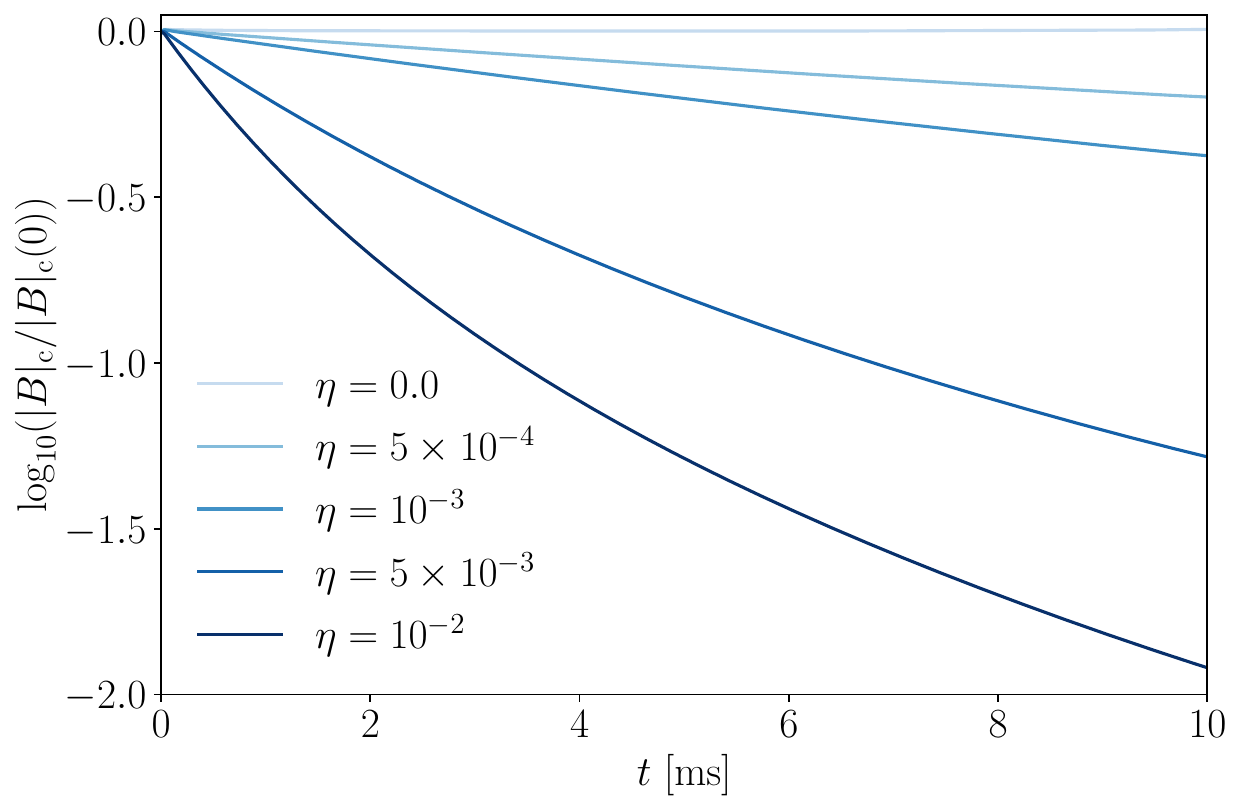}
    \caption{Time evolution of the central magnetic field strength (normalised by its initial value), for resistivities ranging from $\eta = 0.0$ to $\eta =10^{-2}$, using a resolution of $\Delta x = 0.2$.}
    \label{fig:bc_tev}
\end{figure}

We show in~\cref{fig:bc_tev} the time evolution of the central magnetic field amplitude (normalised by its initial value), for different values of the resistivity. As expected, larger values of resistivity lead to a more rapid decrease of the central magnetic field amplitude, due to the Ohmic diffusion introduced by a physical finite conductivity. The decay is exponential, although it looks super exponential for very high resistivities (see the case with $\eta = 10^{-2}$). Fluid velocities are negligible in these simulations, which means that the magnetic field decay is described as a simple diffusive process, as also seen in~\cite{Dionysopoulou:2013,Azizi:2026,BAMpaper}.

\begin{figure*}[htbp]
    \centering
    \includegraphics[width=0.85\textwidth]{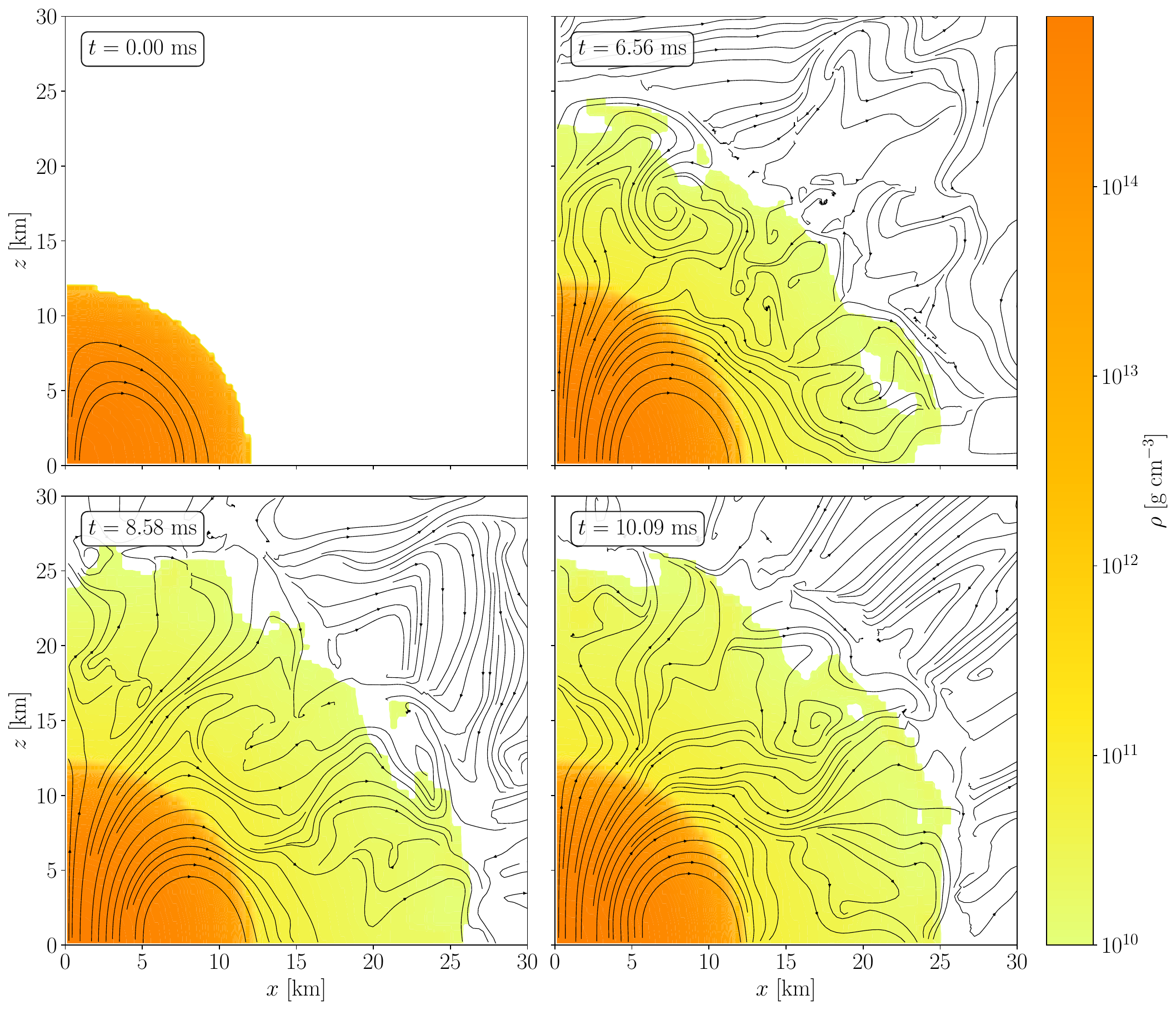}
    \caption{Two-dimensional slices in the $xz$-plane of the rest-mass density, $\rho$, and magnetic field lines, with $\Delta x = 0.2$ and $\eta = 10^{-2}$. Each panel illustrates a different time. At $t = 0$ (top left panel) the magnetic field is confined inside the star, but as the star evolves in time, the magnetic field diffuses towards the exterior. The magnetic field structure outside the star does not present a dipolar structure because, in the exterior, the physical resistivity is zero.}
    \label{fig:rho_field_lines}
\end{figure*}

In~\cref{fig:rho_field_lines}, we plot the mass density contours in the $xz$-plane, together with the magnetic field lines, at different times. Since we do not assume vacuum in the exterior of the star, the magnetic field lines do not rearrange in a dipolar-like shape on the timescale of the simulation. In contrast, the zero-resistivity exterior gives us a very messy magnetic field. With time, the star expands, and regions with $\rho \sim \qty{e10}{\g\per\cm\cubed}$ reach a radius of $\approx \qty{25}{\km}$. As the helicity dissipation rate is linked to the resistivity, we should expect (compared to IMEX methods which typically use a resistivity profile that diverges in the exterior) the evolution to take much longer to order the magnetic field.

The results presented in this section demonstrate that REGIME captures the expected resistive behaviour of a magnetised neutron star, with finite resistivity leading to the diffusion and decay of the magnetic field. Together with the good behaviour in the ideal limit, these results provide a validation of the implementation before considering the more dynamical binary neutron star merger simulations in the following section.

\section{Binary neutron star merger simulations}\label{sec::bns_simulations}

We have performed GRMHD simulations of binary neutron star mergers to investigate the impact of finite resistivity on the post-merger dynamics and associated observable signatures. Binary neutron star mergers provide a particularly relevant test case for the REGIME framework, as they involve strongly magnetised, highly conducting matter and complex magnetic-field evolution on dynamical timescales. We compare simulations performed in the ideal-MHD and finite-resistivity regimes, focusing on the gravitational-wave signal, the structure and evolution of the post-merger remnant, and the amplification and redistribution of the magnetic field. These simulations provide a proof-of-principle demonstration of the application of REGIME to a fully dynamical astrophysical system and allow us to assess whether small departures from the ideal-MHD limit can produce appreciable differences in the post-merger evolution.

\subsection{Initial data}

The initial data is an equal-mass binary neutron star system, constituted by two identical irrotational NSs, modelled using a 7-piece piecewise polytropic representation of the Skyrme-Lyon nuclear EOS~\cite{Chabanat:1998}, which broadly aligns with the current observational constraints of NSs masses and radii~\cite{Riley:2021,Fonseca:2016,Antoniadis:2013,Cromartie:2020,Pang:2021,LVK:2018}. We add an ideal-gas-like term to account for thermal effects, with $\Gamma_\mathrm{th} = 1.8$~\cite{Hotokezaka:2013}. The baryonic mass of each NS is $M_\mathrm{bar} = 1.4$ M$_\odot$ (resulting in a total ADM mass of $\qty{2.53}{\Mass\Sun}$ for the binary), with an equatorial radius of $R_\mathrm{eq} = \qty{9.58}{\km}$. The initial orbital separation is \qty{35}{\km}, with an orbital frequency of \qty{390.35}{\Hz}. The initial data are computed using the \texttt{LORENE} code~\footnote{\url{https://gitlab.in2p3.fr/lorene/Lorene}}~\cite{LORENE:2016}. We set the atmosphere at a baryonic mass density of $\rho_\mathrm{atm} = \qty{6.173e5}{\g\per\cm\cubed}$.

The initial magnetic field is added through the thorn \texttt{SeedMagneticFields}, with a purely poloidal structure, and confined inside the star. The vector potential that leads to this initial magnetic field configuration has a non-vanishing $A_\phi$ component, given by~\cref{eq:Aphi}, with $n_s = 2$, and $P_\mathrm{cut} = \num{4.55e-7}$ ($4\%$ of the maximum pressure). The initial magnetic field strength is $B_0 = \qty{e15}{\gauss}$, which is large enough to overcome the turbulent magnetic field amplification due to the Kelvin-Helmholtz instability during merger. At the employed resolution, the instability and amplification  cannot be fully captured with lower magnetic fields. This choice of initial magnetic field strength is likely larger than the expected large-scale magnetic field strengths of neutron stars, but has been widely used in recent works~\cite{Ruiz:2016, Ciolfi:2017, Combi:2023,DeHaas:2024,Most:2023a,Most:2023b,Kiuchi:2023,Kiuchi:2024}. For this work, we choose a value of $\eta = 10^{-6}$ for the resistivity, equivalent to a conductivity of $\sigma = \qty{2e11}{\per\s}$. This choice is motivated by previous works, such as~\cite{Dionysopoulou:2015,BAMpaper}.

\subsection{Grid structure}

The grid structure consists of two nested refinement boxes, each centred on each NS. Each set contains 8 refinement levels, with resolutions scaling by a factor of two. When merger occurs and the two grids overlap, they merge into a single box centred around the centre of mass of the system. The finest level surrounding each star has a half-length approximately 1.25 times larger than the equatorial radius of the star. The full domain of the simulation is $\left(-1019.2, 1019.2 \right)$ for all three Cartesian directions. In these simulations, the grid spacing of the most refined level is \qty{156.8}{\m}. In a follow-up work, we will discuss both the grid convergence and also the physics impact of the resistivity.

\subsection{Results}

\begin{figure*}[htbp]
    \centering
    \includegraphics[width=\textwidth]{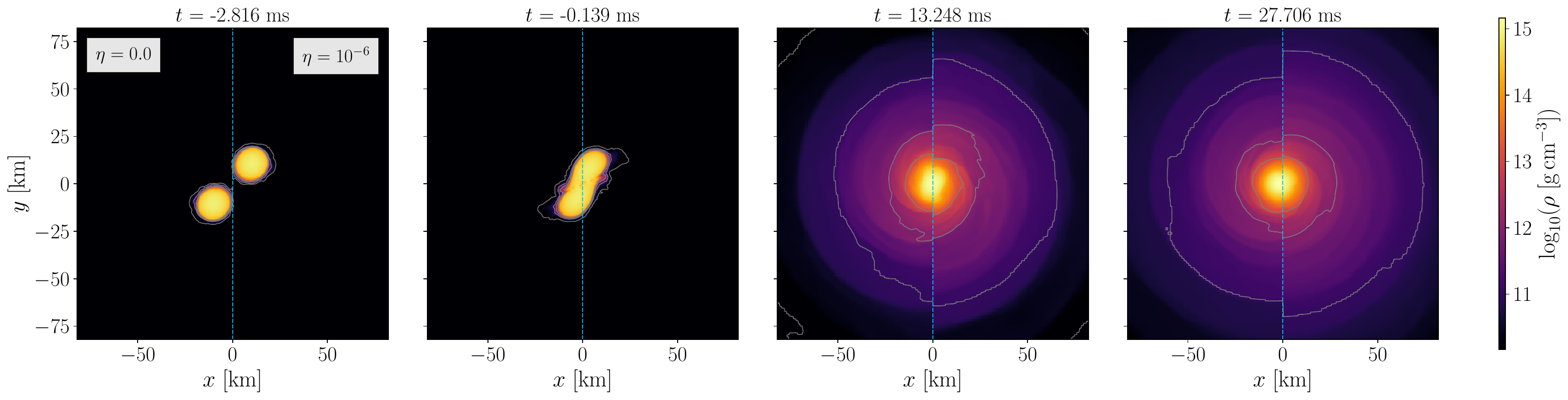}
    \caption{Snapshots of the density in the equatorial $xy$-plane for the binary neutron star merger. The four times shown roughly represent late inspiral, merger, and post-merger phases once the remnant has settled down. The left half of each plot shows the ideal MHD case. The right half of each panel shows the resistive case. The differences are negligible through merger and remain small in the later plots. However, there is a noticeable change in the compactness of the remnant, most easily seen through the density contours, depicted with grey lines.}
    \label{fig:bns_rho_contours_xy}
\end{figure*}

\Cref{fig:bns_rho_contours_xy} depicts the baryonic mass density contours in the equatorial plane at different times (where $t= 0$ corresponds to the time at which the GW amplitude is maximum, i.e.\ merger time). The left half of each panel shows the ideal case, whereas the right side stands for the simulation with $\eta = \num{e-6}$. Before merger, we do not see any noticeable differences. However, at late times after merger (which happens at around $t\approx \qty{5}{\ms}$ after the start of the simulation), the extension of the remnant disc seems to be larger in the resistive case. 

\subsubsection{Gravitational-wave signal}

\begin{figure*}[htbp]
    \centering
    \includegraphics[width=0.6\textwidth]{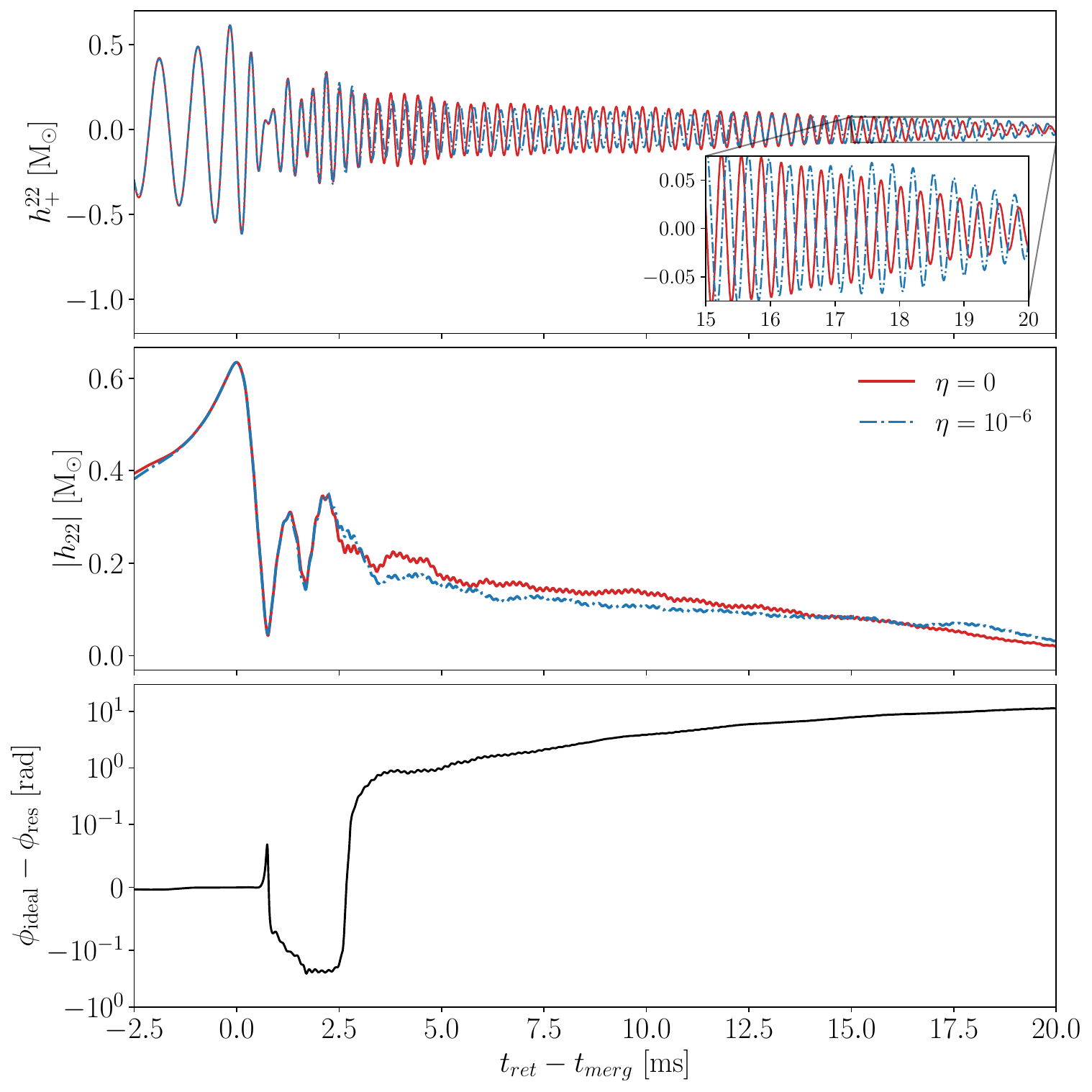}
    \caption{The GW strain comparison of the ideal and resistive cases (top panel). As with the density snapshots, no differences are visible until a few \unit{\ms} after merger. There exist differences in both amplitude and phase (middle and bottom panels, respectively). The likely cause is the change in the compactness of the remnant, as illustrated by~\cref{fig:bns_rho_contours_xy}.}
    \label{fig:bns_hplus}
\end{figure*}

Differences in the GW signal can be seen in~\cref{fig:bns_hplus}. The top panel depicts the (2,2)-mode contribution to the ``+'' component of the GW strain. In red, we plot the ideal case, which is similar to the resistive case (blue dash-dotted line) for the timescale of the simulation. In the middle panel we plot the amplitude envelope of the signal. The lower panel shows the difference between the cumulative phases of both signals, aligned at merger time. The phase difference shows more interesting features than the amplitude of the signal. Right after merger, the phase difference starts having negative values, reaching an amplitude of $\sim 0.1$ rad. After $t\approx \qty{2.5}{ms}$ post-merger, the phase difference starts increasing monotonically, until it reaches a value of more than 10 rad. This effect is likely linked to changes in the bulk dynamics. 

\subsubsection{Compactness}

\begin{figure}[htbp]
    \centering
    \includegraphics[width=0.45\textwidth]{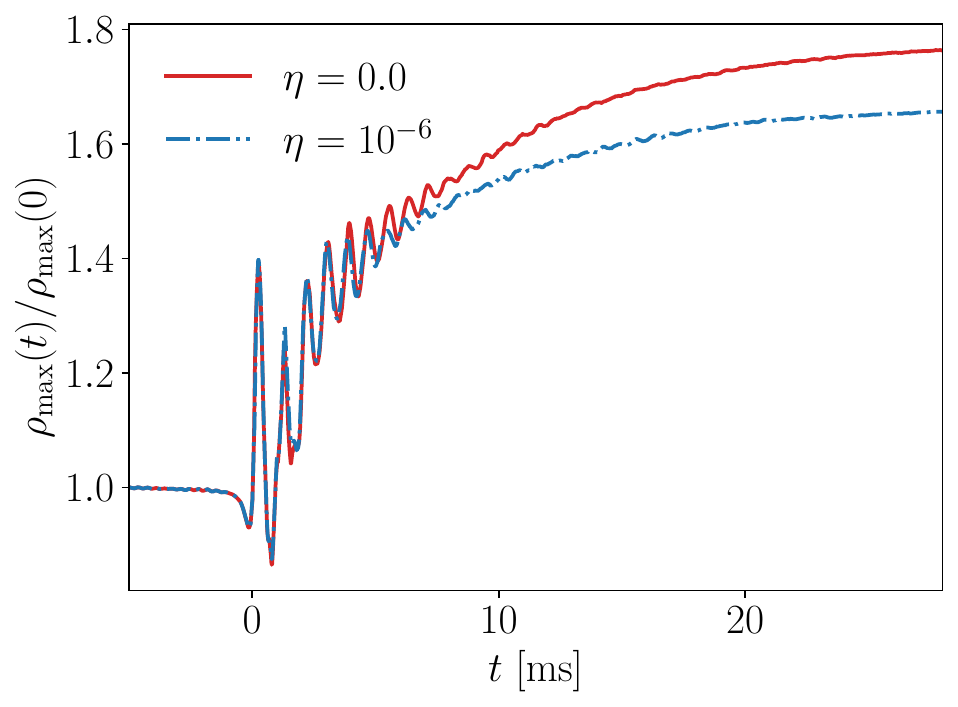}
    \caption{The maximum density (normalized to its initial value) as a function of time. This shows the expected increase through merger and as matter accretes onto the remnant. The resistive case shows slower accretion at late times. This is again linked to the change in compactness in this case. However, the main deviations between the resistive and ideal cases appear later in the maximum density than in observables such as the GW signal. This is likely explained by the differences resulting from the magnetic field, which peaks away from the centre.}
    \label{fig:bns_rho_max}
\end{figure}

The differences in the GW emission are plausibly associated with changes in the compactness of the merger remnant. Regarding the central regions of the object, in~\cref{fig:bns_rho_max} one can see that the maximum baryonic mass density starts deviating at $t\approx \qty{7.5}{\ms}$ after merger. It becomes noticeably larger for the ideal case, indicating that the remnant becomes more compact in the ideal simulation. These differences appear at slightly later times than those observed in the GW signal. This may be related to the magnetic-field structure, whose maximum amplitude is reached away from the centre, allowing differences at finite radii to develop before becoming apparent in the central density.

\begin{figure*}[htbp]
    \centering
    \includegraphics[width=\textwidth]{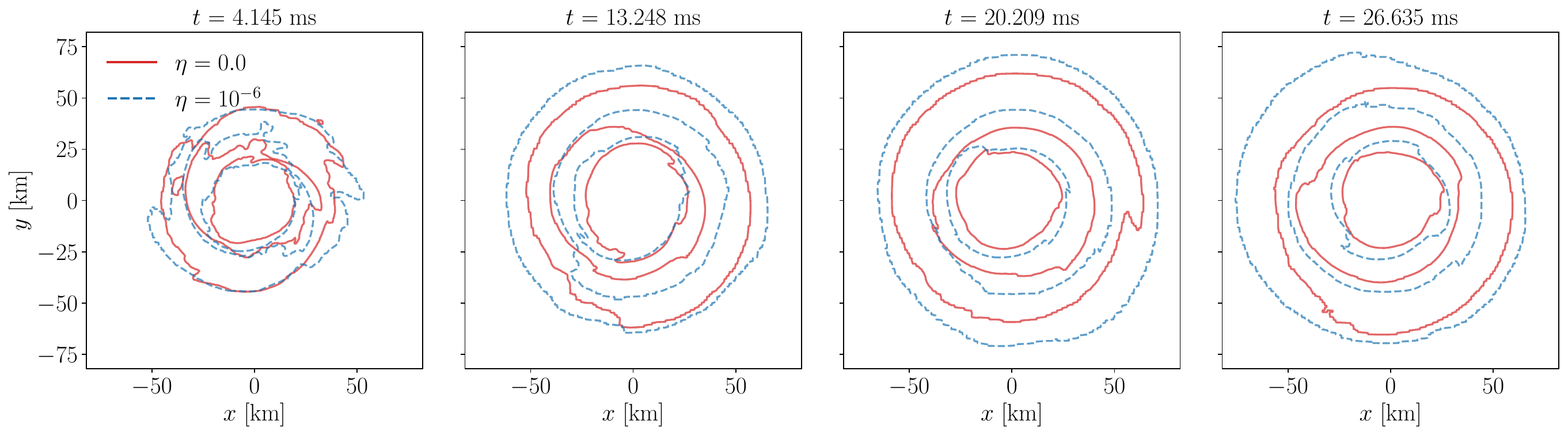}
    \includegraphics[width=\textwidth]{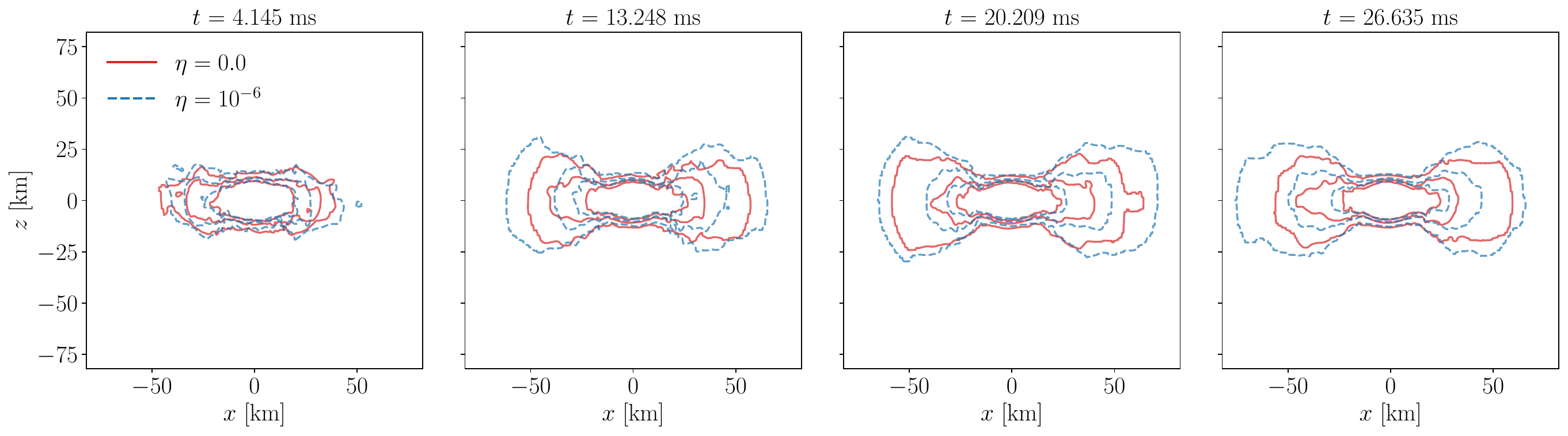}
    \caption{A comparison of representative density contours in the remnant (at densities \(\left\{10^{12},\,10^{11.5},\,10^{11}\right\}\,\unit{\g\per\cm\cubed}\)), showing the ideal results in red and the resistive results in blue. The top row shows cuts through the equatorial $xy$-plane, and the bottom row cuts in the $xz$-plane to show the disc structure. The change in compactness, and that it increases with time, is visible.}
    \label{fig:bns_rho_levels}
\end{figure*}

In~\cref{fig:bns_rho_levels}, we show mass density contour levels in the equatorial plane (top row) and in the $xz$-plane at $y=0$. We set contour levels at \(\left\{10^{12},\,10^{11.5},\,10^{11}\right\}\,\unit{\g\per\cm\cubed}\). At early times after merger (see left-most panels), the disc extension is similar in both simulations. From $t \approx 12$ ms onwards, however, all density levels reach larger radii in the resistive case. The $xz$-plane cuts show that this difference extends over the vertical structure of the disc. Thus, the resistive simulation develops a less compact central remnant and a more extended disc, whereas the ideal simulation remains more compact and has a less extended disc.

\subsubsection{Magnetic field strength and structure}

\begin{figure*}[htbp]
    \centering
    \includegraphics[width=0.8\textwidth]{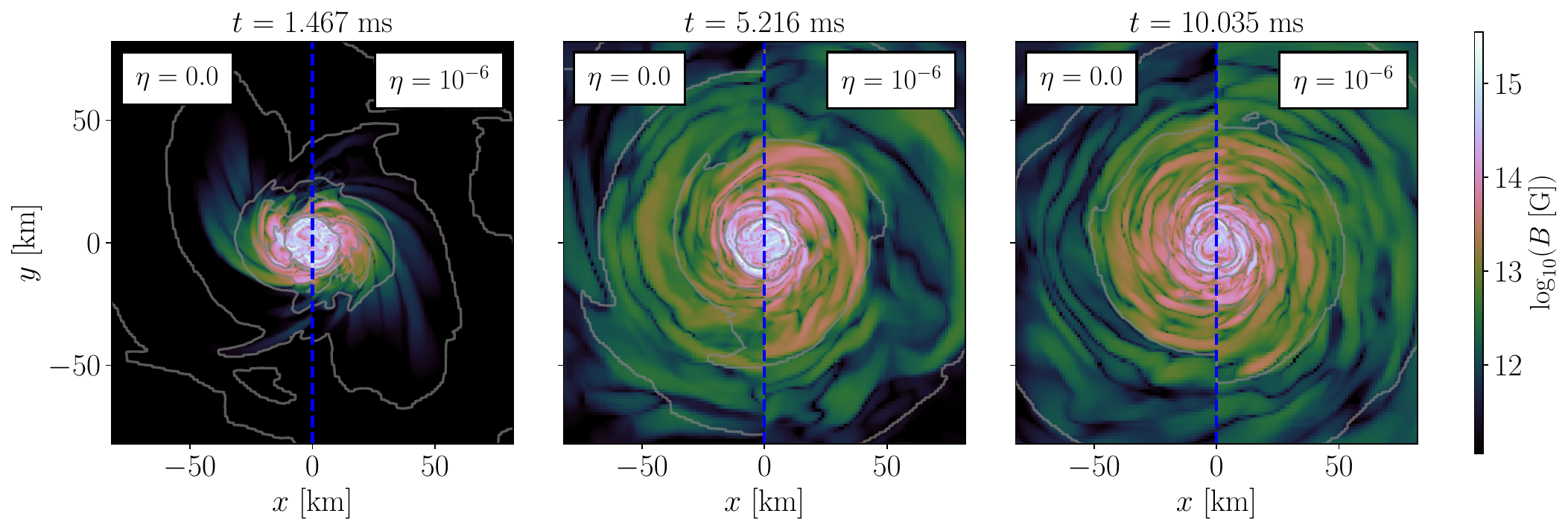}
    \includegraphics[width=0.8\textwidth]{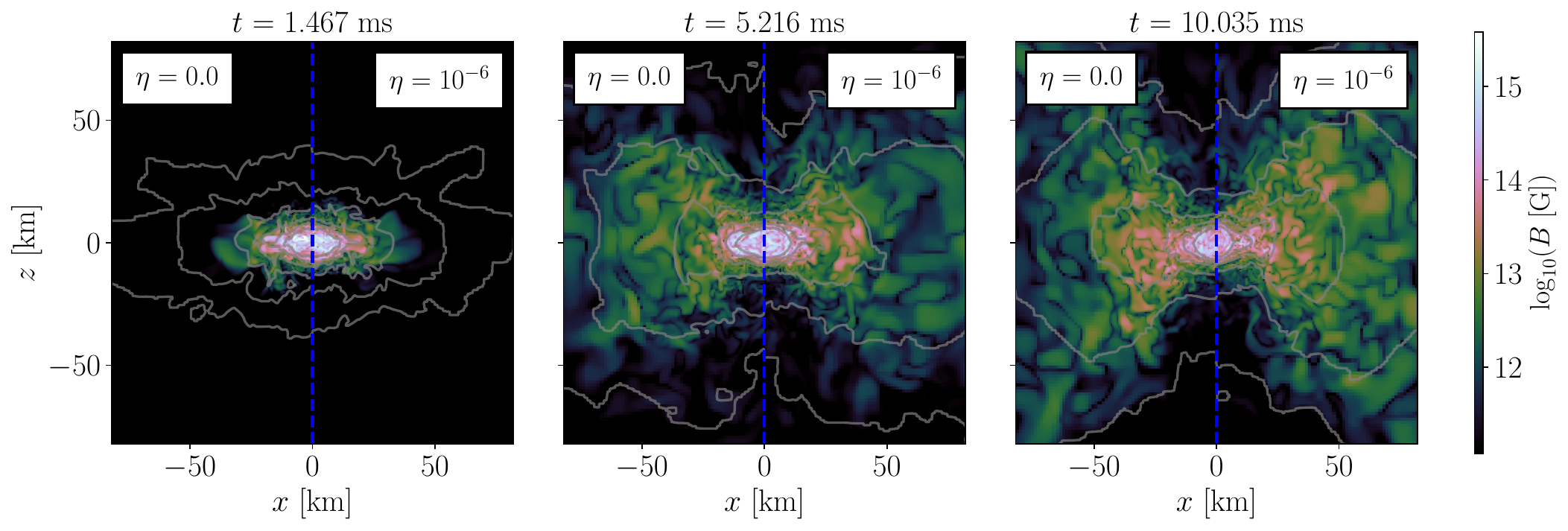}
    \caption{The magnetic field amplitude in the equatorial $xy$-plane (top) and in the $xz$-plane (bottom), showing the different spatial distribution of the magnetic-field amplitude between the ideal and resistive simulations, at early times after merger. The resistive simulation exhibits a larger amplitude, especially in the outermost regions of the remnant.}
    \label{fig:bns_b_ampl_early}
\end{figure*}

\Cref{fig:bns_b_ampl_early} shows two-dimensional contours of the magnetic field amplitude in the equatorial $xy$-plane (top row) and the $xz$-plane (bottom row), with density contours overlaid in grey, at early times after merger. As discussed above, the magnetic field reaches its largest amplitudes away from the centre (see the second column), which may help explain the delayed changes in the central mass density. A few milliseconds after merger, the field amplitude in the $xy$-plane (top panels) reaches $\sim \qty{1e14}{\gauss}$ at radial distances of up to $r\approx \qty{30}{\km}$ in the resistive simulation. By contrast, in the ideal simulation, it decreases to $\sim \qty{5e13}{\gauss}$ at $r\approx \qty{25}{\km}$. In the third column, at $t = \qty{10.035}{\ms}$, the field amplitude in the ideal simulation has decreased to $\sim \qty{1e12}{\gauss}$ at distances $r\gtrsim \qty{50}{\km}$, whereas it remains close to $\qty{1e13}{\gauss}$ in the resistive simulation. The $xz$-plane (bottom panels) exhibits similar behaviour near the equatorial plane. However, at $t = \qty{5.216}{\ms}$ (second column), the field amplitude is slightly larger in the ideal simulation around $|z|\lesssim \qty{50}{\km}$ and $|x|\approx \qty{50}{\km}$. By the time shown in the third column, the field has decayed in the ideal simulation but has increased in the resistive case at locations away from the centre. In both simulations, the field amplitude decreases in the innermost region as the remnant evolves, as seen in both planes.

\begin{figure*}[htbp]
    \centering
    \includegraphics[width=0.8\textwidth]{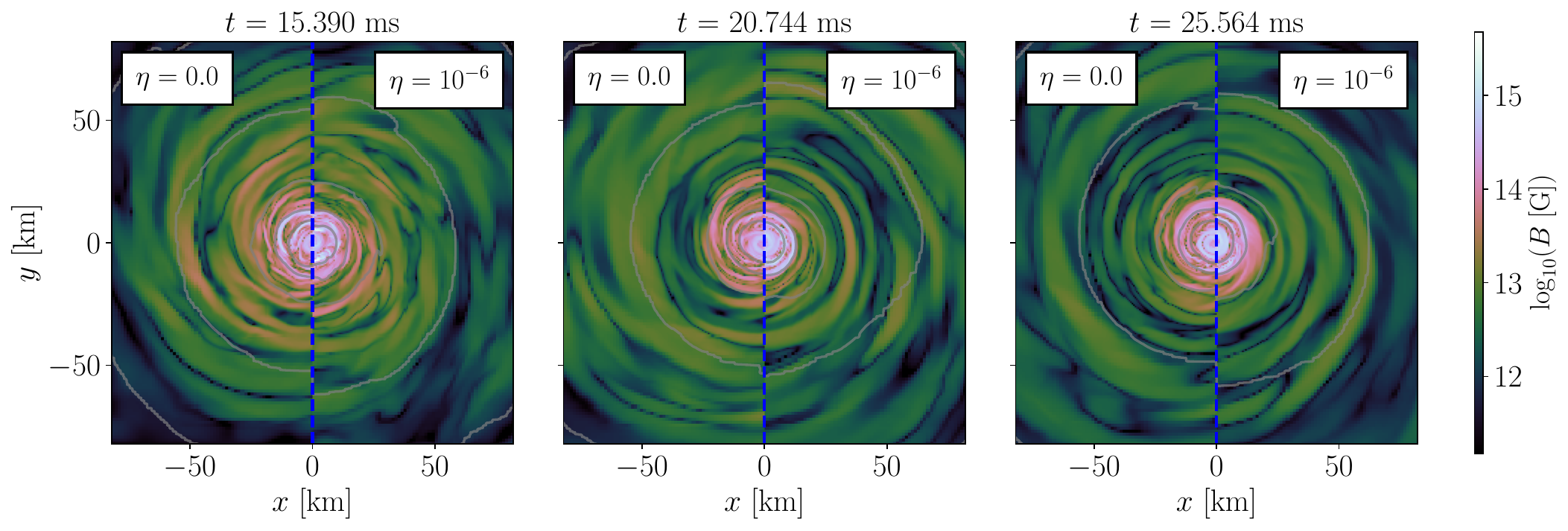}
    \includegraphics[width=0.8\textwidth]{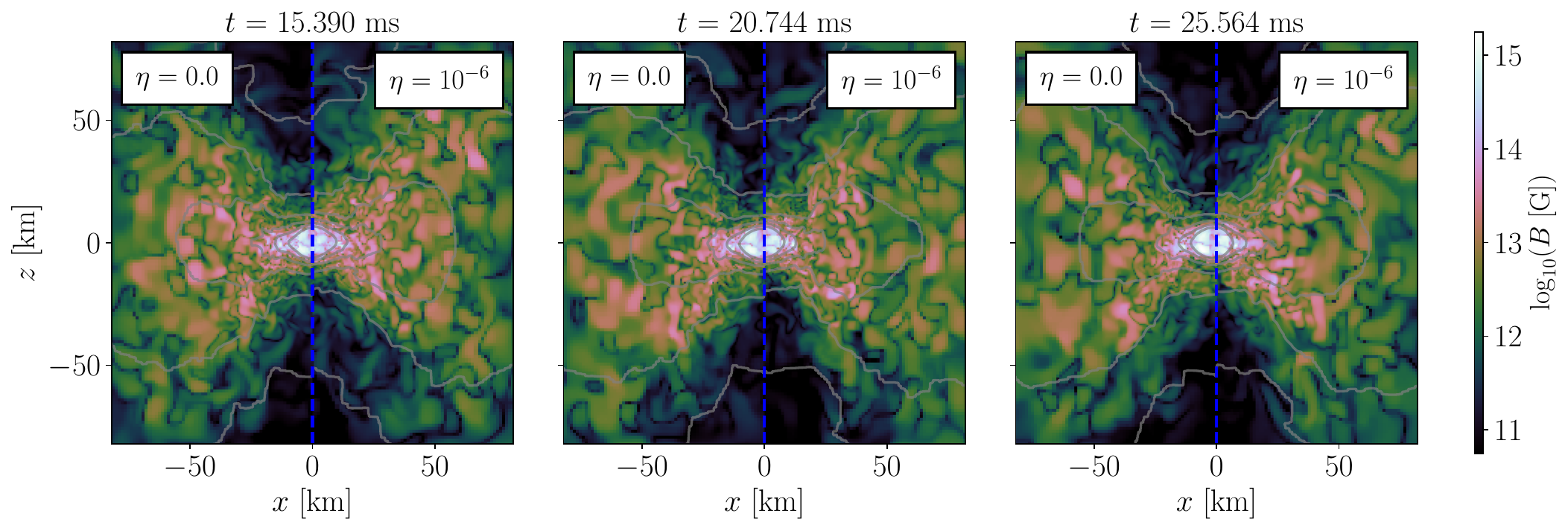}
    \caption{The magnetic field amplitude in the equatorial $xy$-plane (top) and in the $xz$-plane (bottom), showing the different spatial distribution of the magnetic-field amplitude between the ideal and resistive simulations, at late times after merger. The resistive simulation now shows a lower amplitude in the low-density outer regions.}
    \label{fig:bns_b_ampl_late}
\end{figure*}

\Cref{fig:bns_b_ampl_late} shows the same magnetic field contours, but at later times. Within the densest regions of the remnant, extending to radii of $\sim \qty{10}{\km}$, the field amplitude is larger in the resistive simulation. Farther out, however, the field is generally stronger in the ideal simulation, although its amplitude remains small compared with that in the central region. At $t = \qty{15.390}{\ms}$ (first column), the field amplitude in the ideal simulation reaches $\sim \qty{1e14}{\gauss}$ at $r\approx \qty{25}{\km}$ in the equatorial plane (top panel), whereas the corresponding amplitude in the resistive simulation is almost an order of magnitude lower. At later times (second and third columns), the field remains stronger in the outermost regions of the remnant in the ideal simulation, while it decays in the resistive case. In the inner remnant, by contrast, the field appears to grow in both simulations. The $xz$-plane (bottom panels) also reveals a few localised regions of enhanced field amplitude near the equator, around $x\approx \qty{10}{\km}$. Away from the centre but still relatively close to the equatorial plane ($|z|\lesssim \qty{50}{\km}$), the field amplitude is slightly larger in the ideal simulation.

\begin{figure}[htbp]
    \centering
    \includegraphics[width=0.45\textwidth]{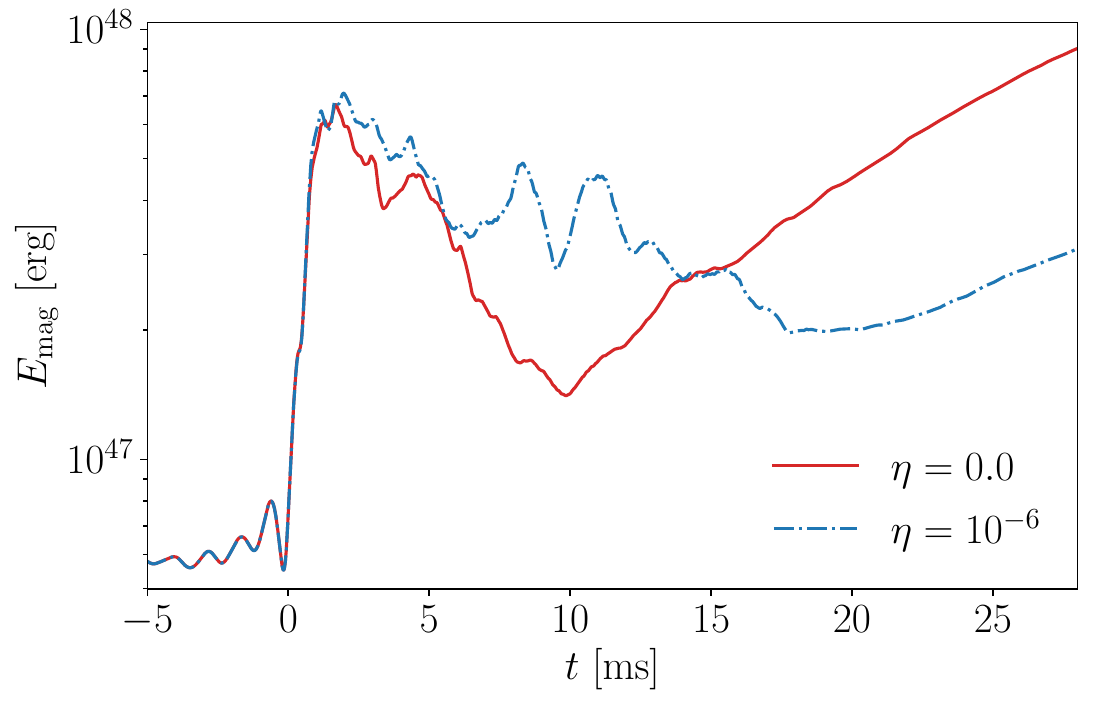}
    \caption{The total magnetic energy as a function of time. The behaviour is similar through the merger, where the energy rapidly increases due to the Kelvin-Helmholtz-like shearing instability in the slip region, with both simulations reaching very similar maximum values. At late times, there is another growth, possibly associated with the development of magnetic instabilities. The differences at intermediate times are substantial: the magnetic energy grows earlier and faster in the ideal case, indicating that the instability sets at earlier times.}
    \label{fig:bns_mag_energy}
\end{figure}

Taken together, \Cref{fig:bns_b_ampl_early,fig:bns_b_ampl_late} suggest a delay in the evolution of the magnetic field between the ideal and resistive simulations. In the ideal case, the field decays more rapidly during the early post-merger phase but begins to grow again sooner than in the resistive simulation. This difference in timing, together with the contrasting field distributions in the inner and outer remnant, indicates that finite resistivity affects not only the magnetic field amplitude but also its spatial and temporal evolution. The evolution of the magnetic field amplitude can also explain the time evolution of the total magnetic energy:
\begin{equation}\label{eq:mag_en}
    E_{\rm mag} = \frac{1}{2}\int_Vb^2\sqrt{\gamma}\,dV\,,
\end{equation}
where $V$ in this case corresponds to the whole simulation domain.~\Cref{fig:bns_mag_energy} shows that, following a rapid increase in magnetic energy immediately after merger, the resistive and ideal simulations reach the same peak amplitude, associated with the development of the Kelvin--Helmholtz instability in the shearing region between the two stars. The magnetic energy subsequently decreases in both simulations at a different rate, being larger in the ideal simulation. In the resistive case, however, the magnetic energy remains approximately constant at $\approx \qty{3.5e47}{erg}$ between $5$ and $12$ ms after merger, consistent with a different redistribution of magnetic energy during this phase. In the ideal case, the magnetic energy begins to grow again at $t\approx 10$ ms post-merger, with a relatively small growth rate, while in the resistive case this subsequent growth starts later, at $t\approx 17$ ms, and proceeds at a  smaller rate. This late-time growth may be associated with the development of magnetic instabilities in the differentially rotating remnant, potentially including the magnetorotational instability. Establishing this connection, however, requires higher spatial resolutions and a dedicated analysis of the instability.

\begin{figure*}[htbp]
    \centering
    \includegraphics[width=0.6\textwidth]{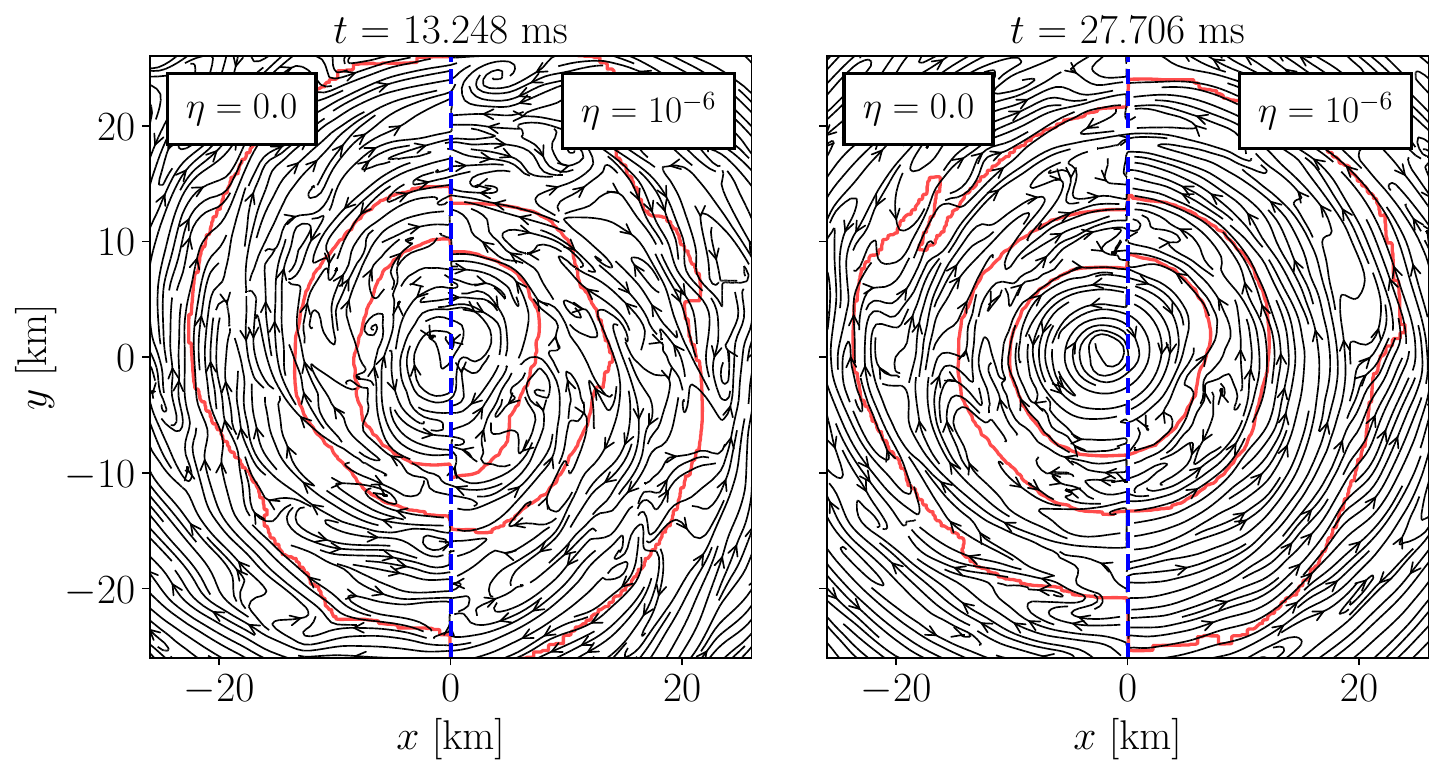}
    \caption{The magnetic field lines in the equatorial $xy$-plane. The red curves depict mass density contours. Particularly at late times (right panel), the additional symmetry and ordering in the resistive case is visually suggested.}
    \label{fig:bns_bfield_lines}
\end{figure*}

\Cref{fig:bns_bfield_lines} shows that the magnetic field is less turbulent in the resistive simulation. The magnetic field lines in the equatorial $xy$-plane show that the field has a toroidal structure. In the resistive case (right half of each panel), the magnetic field lines seem to be more ordered, especially at later times (right panel), supporting the idea that the formation of turbulent structures is suppressed and suggesting that resistivity promotes the reorganisation of the field into larger-scale coherent structures. 

The more ordered magnetic-field structure observed in the resistive simulation, together with the lower magnetic-field amplitude in the inner regions of the remnant, may contribute to the differences in the remnant and disc structure between the ideal and resistive cases. In particular, the suppression of small-scale magnetic structures and the weaker magnetic field in the outer regions may modify the Maxwell stresses responsible for angular-momentum transport, potentially leading to a less efficient extraction of angular momentum from the central region of the remnant. The core would then retain more angular momentum and rotational support, contracting less, and the material would remain at larger radii, resulting in a more extended disc. This interpretation is consistent with the less compact remnant and more extended disc observed in the resistive case, although a direct analysis of the angular-momentum transport is required to establish this connection. Such an analysis is beyond the scope of the present work and will be addressed in a future study. 

\section{Discussion and conclusions}\label{sec::conclusions}

We have presented REGIME, a new framework to introduce finite resistivity in GRMHD simulations. This is the general-relativistic extension of the special-relativistic approach developed in~\cite{Wright:2020}, and a simpler version than that presented in~\cite{Wright_thesis}. Standard approaches typically rely on IMEX schemes to handle the stiff source terms arising from finite resistivity in the electromagnetic equations. While these methods provide a robust treatment of the stiff relaxation associated with high conductivities, their implementation requires substantial modifications to existing GRMHD codes, including the introduction of new evolved fields, updates to the conserved-to-primitive recovery procedure, and the implementation of new time integrators. Moreover, the implicit treatment generally requires nonlinear solves at each timestep, increasing both the computational cost and the complexity of the numerical infrastructure.

The REGIME framework provides an alternative perturbative approach for approximating resistive GRMHD in the high-conductivity regime. Using a perturbative expansion around the ideal-MHD limit~\cite{Wright:2020,Hatton:2024}, REGIME captures the leading non-ideal effects without requiring implicit time integration. Instead, it introduces a correction term in the evolution equation for the vector potential that represents the first-order resistive contribution around the ideal-MHD equilibrium state. This correction is proportional to the inverse conductivity, which is assumed to be small, as expected in many astrophysical scenarios such as neutron star mergers~\cite{Harutyunyan:2018}. Moreover, this framework is alternative to the one developed in~\cite{Wright_thesis}. In this work, we use fluid-frame variables to exploit algebraic relations, avoiding the complex matrix inversions associated with the use of Eulerian variables.

The numerical implementation of this scheme has been done within the \texttt{IllinoisGRMHD} code~\cite{IllinoisGRMHD:2015}, built on the \texttt{GRHayL} library~\cite{Cupp:2026}, and part of the Einstein Toolkit infrastructure~\cite{Loffler:2012}. Since the vector potential in \texttt{IllinoisGRMHD} lives in the edges of the numerical cells, the implementation of REGIME requires interpolating the velocity and magnetic fields, together with the spacetime variables, to faces (for the second order centred finite differences used to compute the spatial derivatives) and edges using a two-point averaging. The simple discretisations performed in this proof-of-principle paper therefore have limited, second-order, numerical accuracy.

Our framework has been validated with several numerical tests. One-dimensional special-relativistic current sheet tests~\cite{Komissarov:2007} allowed us to compare the numerical solution of the diffusion equation for the magnetic field with its known analytical result. The robustness of the REGIME approach has been verified, and we have observed that the framework provides second order accuracy, as also seen in other works that employ IMEX schemes~\cite{Dionysopoulou:2013,Cheong:2022,Azizi:2026,BAMpaper}.

We have also performed fully general-relativistic tests, evolving a non-rotating magnetised TOV star in a fully dynamical spacetime. The magnetic field has been inserted in the neutron star so that it is confined to its interior. The resistivity has been modelled using a simple profile that varies spatially as a function of the baryonic mass density. The profile smoothly transitions between a fully resistive regime in the stellar interior to (near) ideal MHD in the exterior. Our results show that the magnetic field dissipates with time, as expected in a resistive regime, and in agreement with previous works~\cite{Dionysopoulou:2013, Azizi:2026}. We do, however, observe differences in the magnetic field structure at low densities compared to these studies, as they assume perfect electrovacuum in the stellar exterior, as opposed to the ideal MHD atmosphere modelled here. 

Our last test of this proof-of-principle paper considers the evolution of a magnetised neutron star binary system. We have considered an equal-mass system of two irrotational neutron stars described by a hybrid piecewise polytropic EOS (SLy), and we have inserted a purely poloidal magnetic field confined within the stars. We have evolved the system up to $\sim \qty{30}{\ms}$ after merger, which is sufficient to study the impact of resistivity on the GW signal and the post-merger remnant structure. We find differences in both the amplitude of the GW signal and its cumulative phase, which develop only a few milliseconds after merger. These differences are accompanied by changes in the post-merger remnant and disc structure: the resistive simulation develops a more extended disc and a less compact central remnant, with a lower maximum baryonic density than in the ideal case.

The differences in the post-merger structure are accompanied by a different evolution and spatial distribution of the magnetic field. At late times, the resistive simulation develops a more ordered large-scale magnetic field, with a lower magnetic-field amplitude in the outermost regions of the remnant and also a lower total magnetic energy. The relaxation of the ideal flux-freezing condition therefore leads not only to magnetic dissipation, but also to a reorganisation of the magnetic-field structure. This reorganisation may modify the Maxwell stresses responsible for angular-momentum transport, potentially leading to less efficient extraction of angular momentum from the central region of the remnant and, consequently, to a less compact remnant and a more extended disc than in the ideal case. A direct analysis of the angular-momentum transport will be required to establish this connection.

To assess the computational cost of the resistive implementation, we compare its performance with that of the standard ideal-MHD version of \texttt{IllinoisGRMHD} distributed with the Einstein Toolkit ``Martin D. Kruskal" release~\cite{2025zndo..15520463R}. At $t \approx \qty{10}{\ms}$ after merger, the ideal-MHD simulation runs approximately 15\% faster than the resistive simulation, indicating a relatively modest computational overhead for the inclusion of finite resistivity. This should not be regarded as the optimal performance of the present implementation, which still contains several computationally expensive operations that could be further optimised. For example, the interpolation operations required to consistently apply the resistive corrections at cell edges could potentially be avoided in a finite-element implementation. More importantly, unlike IMEX schemes, whose computational cost generally increases as the resistive timescale becomes shorter and the conductivity increases, the REGIME formulation does not introduce a corresponding stiffness-driven increase in computational cost. The overhead associated with finite resistivity can therefore remain moderate even as the system approaches the ideal-MHD limit.

The results presented in this work suggest that finite resistivity can influence the post-merger evolution through changes in the magnetic-field structure and its associated stresses, potentially affecting the redistribution of mass and angular momentum and the subsequent formation of magnetically driven outflows. Establishing a quantitative connection between resistivity and these effects will require higher-resolution simulations, longer post-merger evolutions, and a dedicated analysis of angular-momentum transport, magnetic-energy spectra, ejecta, and outflow evolution.

\begin{acknowledgments}
The authors would like to thank Nils Andersson for useful discussions and for providing insightful comments on the manuscript. MMT and IH gratefully acknowledge support from the Science and Technology Facilities Council (STFC) via grant number ST/Y000811/1. MMT was additionally supported by The Computational Science Centre for Research Communities (CoSeC) 2025 Fellowship programme, and by the Spanish Agencia Estatal de Investigación (grant PID2024-159689NB-C21) funded by MICIU/AEI/10.13039/501100011033 and by FEDER/EU. The authors acknowledge the use of the IRIDIS 6 High Performance Computing Facility and associated support services at the University of Southampton. This work used the DiRAC Memory Intensive service (Cosma8) at Durham University, managed by the Institute for Computational Cosmology on behalf of the STFC DiRAC HPC Facility (www.dirac.ac.uk), under DiRAC RAC18 allocation APP91053. The DiRAC service at Durham was funded by BEIS, UKRI and STFC capital funding, Durham University and STFC operations grants. DiRAC is part of the UKRI Digital Research Infrastructure. 

This work has used the following open-source packages: \textsc{NumPy}~\citep{harris:2020}, \textsc{SciPy}~\citep{scipy:2020}, \textsc{h5py}~\citep{h5py2025} and \textsc{Matplotlib}~\citep{Hunter:2007}. Claude by Anthropic and GitHub Copilot have been used to aid in code development. All output was reviewed and verified by the authors, who assume full responsibility for the contents of this paper.
\end{acknowledgments}

\appendix

\section{Alternative derivation}

The construction of the correction term in the REGIME method given in~\cref{sec:regime_eqns} follows the approach used in similar models containing a stiff source that enforces a relaxation to equilibrium. However, the algebraic nature of the assumed form of the electric field, combined with the relaxation state being trivial in the fluid frame, allows an alternative derivation that easily extends to more complex cases.

Start from
\begin{equation}
    F_{ab} = u_a e_b - u_b e_a + \varepsilon_{abcd} u^c b^d
\end{equation}
and
\begin{equation}
    \perp^a_b \nabla_c F^{ab} = \left( \delta^a_b + u^a u_b \right) \nabla_c F^{ab} = j^a \, .
\end{equation}

Note that the electric field in the fluid frame vanishes in the ideal limit. Write
\begin{equation}
    e_a = \epsilon e_a^{(1)}
\end{equation}
where $\epsilon$ controls the approach to the ideal limit. Therefore, the Faraday tensor can be written as
\begin{equation}
    F_{ab} = \varepsilon_{abcd} u^c b^d + \mathcal{O}(\epsilon) \, ,
\end{equation}
leading directly to 
\begin{equation}
    j^a = j^a_{(0)} + \mathcal{O}(\epsilon) =  \perp^a_b \nabla_c \left( \varepsilon^{abcd} u_c b_d \right) + \mathcal{O}(\epsilon) \, .
\end{equation}
Note that the leading order term $j^a_{(0)}$ can be explicitly computed from the standard MHD evolved quantities.

Now, the ansatz for the electric field is chosen to be
\begin{equation}
    e_a = \epsilon e_a^{(1)} = \epsilon \left( \tilde{\eta} j_a + \tilde{\xi} b_a + \tilde{\zeta} \varepsilon_{abcd} j^b b^c u^d \right) \, .
\end{equation}
We note that, as $e_a u^a = 0$ by definition, this is a general form for the leading order correction away from the ideal limit, but is not completely generic (as, for example, there are degenerate cases where $j_a$ and $b_a$ may be parallel). We can therefore substitute in $j^a_{(0)}$ explicitly, finding
\begin{equation}
\label{eq:app_regime_full}
    e_a^{(1)} = \tilde{\eta} j^{(0)}_a + \tilde{\xi} b_a + \tilde{\zeta} \varepsilon_{abcd} j_{(0)}^b b^c u^d \, ,
\end{equation}
which can also be explicitly computed.

The first term in~\cref{eq:app_regime_full} is exactly the REGIME resistivity term. Its practical implementation follows the steps laid out in the body of the paper. However, the other terms also have interesting physical interpretations. The second term proportional to $\tilde{\xi}$ is referred to by~\cite{Bucciantini:2013} as a dynamo term. By following through the steps to convert to the spatial frame, it can be qualitatively compared to, for example, \cite{ngSpinningNeutronstarMerger2026} where a phenomenological model for a Taylor-Spruit dynamo is employed. Similarly, the final term can be linked to the Hall drift. This is because, in the Newtonian limit, $j_{(0)}^b \to \mathbf{v} \times \mathbf{B}$ and so the term proportional to $\tilde{\zeta}$ ``is'' the Hall drift term.

\bibliographystyle{apsrev4-1}
\bibliography{sample}

\end{document}